\documentclass{article}

\usepackage[british]{babel}

\usepackage[letterpaper,top=2cm,bottom=2cm,left=3cm,right=3cm,marginparwidth=1.75cm]{geometry}

\usepackage{amsmath}
\usepackage{graphicx}
\usepackage[colorlinks=true, allcolors=blue]{hyperref}
\usepackage[authoryear,round]{natbib}
\usepackage{todonotes}
\usepackage{subcaption}
\usepackage{xcolor}
\usepackage{outlines}
\usepackage{pifont}
\usepackage{authblk}

\title{High-order Simulation of Nearly Hydrostatic Atmospheric Flows}

\author[1]{Shoyon Panday}
\author[1]{St\'ephane Gaudreault}
\author[1]{Carlos Pereira}
\author[2]{Philipp Birken}

\affil[1]{Recherche en Pr\'evision Numérique Atmosph\'erique, Environnement et Changement Climatique Canada, 2121 Route Transcanadienne, Dorval, H9P 1J3, QC, Canada}
\affil[2]{Lund University, Centre for Mathematical Sciences, Lund, Sweden}

\begin{document}
\date{}
\maketitle

\begin{abstract}
Accurate simulation of atmospheric flows is challenging. Low-Mach-number dynamics often need to be resolved in the presence of a dominant, nearly hydrostatic background state. Conventional upwind numerical fluxes can introduce excessive acoustic-speed-based dissipation in the low-Mach regime. Discretization errors in the balance law can also generate spurious motion. This work investigates these two issues within a high-order Direct Flux Reconstruction (DFR) framework coupled with a third-order exponential time integrator for the 3D compressible Euler equations with gravity. Low-Mach-number accuracy is addressed using a modified AUSM$^+$-up flux, denoted AUSM$^+$-upH. Preservation of the background equilibrium is improved through a deviation-based formulation. The resulting approach is assessed using a hierarchy of idealized atmospheric test cases. The results show that AUSM$^+$-upH preserves low-Mach flow structures more effectively than the more dissipative Rusanov flux. However, reduced numerical dissipation also makes AUSM$^+$-upH more sensitive to residual errors in the discrete hydrostatic balance. The deviation formulation substantially suppresses these errors by several orders of magnitude, with the strongest improvement observed for flows over orography. Overall, AUSM$^+$-upH exhibits greater robustness when combined with the deviation formulation across the atmospheric test cases considered.
\end{abstract}

\section{Introduction}

In numerical weather prediction (NWP), high-order methods are receiving increasing attention as alternatives to traditional low-order approaches because of their ability to provide high accuracy, favourable dispersion and dissipation characteristics, and efficient usage of modern high-performance computing architectures \citep{GIRALDO20083849, Marras2016, KELLY20127988}. Atmospheric flows, however, present a challenging numerical setting. Under the action of gravity, the atmosphere is vertically stratified and often evolves close to hydrostatic equilibrium, with relevant dynamics appearing as small perturbations of this background state \citep{BOTTA2004539}. Accurately maintaining this balanced stratified state over long time periods on a large domain is therefore essential. At the same time, atmospheric models must accommodate a wide range of characteristic velocities, from slowly evolving balanced motions to fast advective flows, while the fully compressible equations also support much faster acoustic waves that are generally not of direct meteorological interest. The coexistence of fast and slow atmospheric processes creates significant numerical difficulties for both spatial discretization and time integration. In particular, the numerical dissipation required to stabilize high-order discretizations of nonlinear hyperbolic equations \citep{SCHWARZ2025106874, Persson, Mengaldo2015, YuJan2020, Ullrich2018} must not overwhelm slowly evolving low-Mach-number motions or small perturbations about hydrostatic equilibrium. 

These considerations motivate the present study, which investigates low-Mach and well-balancing techniques within a high order direct flux reconstruction (DFR) framework \citep{romero2016simplified, romero2017development} for the three-dimensional Euler equations with gravity. For the time integration, we use the exponential time integrator EPI3 \citep{TOKMAN2006748}, which is well suited for integrating stiff systems arising from discontinuous Galerkin (DG) like discretizations \citep{birken:21,Clancy01122013}.

Low-Mach-number flows present two major challenges for compressible numerical schemes. First, the large disparity between slow material waves and fast acoustic waves introduces stiffness. Second, classical upwind schemes lose accuracy in the incompressible limit because their numerical dissipation does not scale appropriately, leading to excessive damping. As the Mach number decreases, this artificial diffusion becomes dominant even when the characteristic spatial scales of the physical flow remain unchanged, preventing the discrete scheme from correctly reproducing the incompressible limit \citep{volpe1993, GuillardNkonga2017}. For high order methods, this has been analyzed in \citep{Jung2024, barsukow-2025}. \citet{GOUASMI2022111036} discuss the interplay of low-Mach preconditioning with high-order entropy-stable schemes. 
 
 To address the loss of spatial accuracy, several approaches have been proposed, including flux preconditioning \citep{weiss1995, Turkel1999, Guillard1999OnTB, Miczek2015}, all-speed schemes \citep{LI20085144, LIOU2006137, CORDIER20125685}, and modified velocity jumps \citep{THORNBER20084873, RIEPER20115263, L2ROE}. These approaches generally adjust the numerical dissipation so that the scheme retains the appropriate asymptotic behaviour and recovers the incompressible limit as the Mach number tends to zero.

Furthermore, atmospheric models must accurately preserve equilibrium states. When the physical flow consists of small-amplitude deviations from a balanced background state, the magnitude of these perturbations can become comparable to or smaller than the local truncation error of the discretization \citep{chertock2023}. Consequently, standard numerical methods may struggle to maintain these equilibria over long time periods, as discretization errors can generate spurious departures from the balanced state, particularly at coarse resolution. Numerical methods specifically designed to preserve stationary solutions exactly, or with enhanced accuracy, are commonly referred to as well-balanced methods \citep{Cargo1994UnS}. Over the years, a broad range of well-balanced discretizations has been developed. For a recent review, see \citep{Kappeli2022}.

Recent studies have combined different numerical approaches to address low-Mach accuracy and well-balancing for the Euler equations with gravity. In a finite-volume framework, \citet{Edelmann2021} combined well-balancing with a low-Mach numerical flux. \citet{ligujizh:25} proposed a well-balanced, entropy-stable, and positivity-preserving nodal DG scheme in 2D, but do not discuss the low Mach number issue. \citet{ARTIANO2026115221} developed structure-preserving DG schemes that combine total-energy conservation with well-balancing through a flux differencing approach for the source term in 3D.  

Here, we choose a different approach by combining three complementary numerical strategies within a DFR framework to address the low-Mach, acoustic-stiffness, and hydrostatic-balance issues encountered in compressible atmospheric flows. A modified AUSM$^+$-up flux called AUSM$^+$-upH \citep{Miczek2013} is used because of its improved low-Mach-number accuracy and robustness obtained through appropriate asymptotic scaling. The AUSM$^+$-upH formulation is coupled with an exponential time integrator, which avoids the nonlinear solves associated with fully implicit methods while allowing the stiff acoustic dynamics to be treated efficiently. Finally, the deviation formulation is used to enhance the preservation of the underlying balanced state \citep{Berberich_Chandrashekar_Klingenberg_2021}. The resulting framework is designed to retain weak low-Mach-number perturbations without introducing excessive numerical dissipation or spurious departures from hydrostatic balance. 

The remainder of the paper is organized as follows. Section \ref{equations} introduces the governing equations and thermodynamic formulation. Section \ref{numrmethod} presents the numerical methodology, including the DFR spatial discretization, the exponential time-integration scheme, and the numerical flux functions. Section \ref{DM} describes the deviation formulation. Section \ref{results} presents a series of numerical experiments used to assess the accuracy and robustness of the proposed methods. Finally, Section \ref{summary} summarizes the main conclusions.

\section{Governing Equations} \label{equations}

In Cartesian coordinates, the three-dimensional compressible Euler equations with a source term can be written in conservative form as
\begin{equation}
    \frac{\partial \mathbf{q}}{\partial t} 
    + \frac{\partial}{\partial x} \mathbf{F}_x(\mathbf{q})
    + \frac{\partial}{\partial y} \mathbf{F}_y(\mathbf{q}) 
    + \frac{\partial}{\partial z} \mathbf{F}_z(\mathbf{q})
    = \mathbf{S}(\mathbf{q}),
    \label{eq:pde}
\end{equation}
where the vector of conserved variables is
\begin{equation}
    \mathbf{q}
    =
    \begin{pmatrix}
        \rho \\
        \rho u \\
        \rho v \\
        \rho w \\
        \rho \theta
    \end{pmatrix}.
\end{equation}
Here, $\rho$ denotes the fluid density, $\theta$ is the potential temperature, and $u$, $v$, and $w$ are the velocity components in the $x$, $y$, and $z$ directions, respectively.

The flux vectors in the three spatial directions contain the mass, momentum, and potential-temperature fluxes and are given by
\begin{equation}
    \mathbf{F}_x(\mathbf{q}) = 
    \begin{pmatrix} 
        \rho u \\ 
        \rho u^2 + p \\ 
        \rho u v \\ 
        \rho u w \\ 
        \rho \theta u 
    \end{pmatrix}, \quad
    \mathbf{F}_y(\mathbf{q}) = 
    \begin{pmatrix} 
        \rho v \\ 
        \rho v u \\ 
        \rho v^2 + p \\ 
        \rho v w \\ 
        \rho \theta v 
    \end{pmatrix}, \quad
    \mathbf{F}_z(\mathbf{q}) = 
    \begin{pmatrix} 
        \rho w \\ 
        \rho w u \\ 
        \rho w v \\ 
        \rho w^2 + p \\ 
        \rho \theta w 
    \end{pmatrix}.
\end{equation}

The corresponding source term due to gravity is
\begin{equation}
    \mathbf{S}(\mathbf{q})
    =
    \begin{pmatrix}
        0 \\
        0 \\
        0 \\
        -\rho g \\
        0
    \end{pmatrix},
\end{equation}
where $p$ is the pressure and $g$ is the constant gravitational acceleration acting in the negative $z$-direction.

The system is closed using the equation of state
\begin{equation}
    p = p_0
    \left(
    \frac{\rho R_d \theta}{p_0}
    \right)^{\gamma_d},
\end{equation}
where the potential temperature is defined as
\begin{equation}
    \theta = T\left(\frac{p_0}{p}\right)^{R_d/c_{pd}}.
    \label{potential}
\end{equation}

Here, $T$ is the temperature in Kelvin, $R_d = c_{pd} - c_{vd}$ is the specific gas constant for dry air,
$\gamma_d = c_{pd}/c_{vd}$ is the ratio of specific heats, and
$p_0 = 100{,}000\ \mathrm{Pa}$ is the reference pressure. In this study, we assume a dry, inviscid atmosphere with no diabatic heating.

The use of potential temperature rather than total energy as a prognostic thermodynamic variable follows a common convention in NWP. When required, potential temperature can be related to the temperature through Eq.~\eqref{potential}.

\section{Numerical Method} \label{numrmethod}
\subsection{Direct Flux Reconstruction}

The spatial discretization is performed using the DFR method \citep{romero2016simplified, romero2017development}. DFR belongs to the broader family of flux-reconstruction (FR) schemes and can be viewed as a variant of the weak-form of nodal DG method \cite{huynh2007flux}. The main distinction is that, in the standard FR formulation, the discontinuous flux is corrected using specially chosen correction functions so that a common interface flux is enforced. In DFR, this correction step is written more directly by constructing a single continuous flux polynomial that includes both the interior flux values and the numerical fluxes at the element boundaries.

For clarity, the DFR formulation is summarized for a one-dimensional conservation law, which provides a concise setting for describing the main concept of the method. Details can be found in \citep{gaudreault2026application}.

Consider the one-dimensional conservation law

\begin{equation}
    \frac{\partial \mathbf{q}}{\partial t}
    + \frac{\partial \mathbf{F}_x(\mathbf{q})}{\partial x}
    = 0, 
    \qquad x \in \mathcal{D}.
    \label{eq:conslaw}
\end{equation}

The domain $\mathcal{D}$ is divided into $N_e$ non-overlapping elements.
A local coordinate variable $\xi$ is defined through an affine transformation that
maps each physical element $D_j$ onto the reference element $I_j=[-1,1]$.

A set of $N_s \ge 1$ solution points is then defined within the reference element.
There are several ways to place these points, leading to schemes with different
numerical properties. In the present study, Gau{\ss}-Legendre points are used
because of their favourable numerical properties
\citep{gassner2011comparison, witherden2020nodal}.

Within each element, the solution and flux are approximated using Lagrange
polynomials of degree $N_s-1$ at the $N_s$ solution points:
\begin{align}
    \widehat{\mathbf{q}}_j(\xi,t)
    &=
    \sum_{k=1}^{N_s}
    \widehat{\mathbf{q}}_{j,k}(t)\ell_k(\xi), \\
    \widehat{\mathbf{f}}_j(\xi,t)
    &=
    \sum_{k=1}^{N_s}
    \widehat{\mathbf{f}}_{j,k}(t)\ell_k(\xi),
    \qquad j=1,\ldots,N_e,
    \label{eq:q_f_approx}
\end{align}
where $\ell_k(\xi)$ are the Lagrange basis polynomials associated with the
solution points, and
$\widehat{\mathbf{f}}_{j,k}(t)
= \mathbf{F_x}(\widehat{\mathbf{q}}_{j,k}(t))$.

At element interfaces, the flux generally has distinct left and right traces, so a numerical flux function is required to define a common interface flux. A detailed description of the numerical flux functions used in this study is provided in Section~\ref{flux}.

A continuous flux polynomial $\widetilde{\mathbf{f}}_j(\xi)$ is
constructed in each element by including the two boundary flux values:
\begin{equation}
    \widetilde{\mathbf{f}}_j(\xi)
    =
    \mathbf{f}^{*}_{j,0}\tilde{\ell}_0(\xi)
    +
    \sum_{k=1}^{N_s}
    \widehat{\mathbf{f}}_{j,k}(t)\tilde{\ell}_k(\xi)
    +
    \mathbf{f}^{*}_{j,N_s+1}\tilde{\ell}_{N_s+1}(\xi).
    \label{eq:cont_flux}
\end{equation}
Here, $\mathbf{f}^{*}_{j,0}$ and $\mathbf{f}^{*}_{j,N_s+1}$ denote the numerical
fluxes at the left and right boundaries of element $D_j$, respectively.
The functions $\tilde{\ell}_n(\xi)$ are Lagrange basis polynomials of degree
$N_s+1$, defined using the $N_s$ interior solution points together with the
two boundary flux points.

The derivative of the continuous flux polynomial is therefore
\begin{equation}
    \widetilde{\mathbf{f}}'_j(\xi)
    =
    \mathbf{f}^{*}_{j,0}\tilde{\ell}'_0(\xi)
    +
    \sum_{k=1}^{N_s}
    \widehat{\mathbf{f}}_{j,k}(t)\tilde{\ell}'_k(\xi)
    +
    \mathbf{f}^{*}_{j,N_s+1}\tilde{\ell}'_{N_s+1}(\xi).
    \label{eq:cont_flux_deriv}
\end{equation}

Substituting the discrete approximations of the solution variables and spatial
derivatives into \eqref{eq:conslaw} yields the semi-discrete system of ordinary
differential equations
\begin{equation}
    \frac{d}{dt}\widehat{\mathbf{q}} = \mathcal{F}(\widehat{\mathbf{q}}),
    \label{eq:semi_discrete}
\end{equation}
which is advanced in time using the method of lines.

For multidimensional problems on a regular grid, the one-dimensional basis is extended by a tensor-product construction, allowing the same discretization procedure to be applied independently in each coordinate direction.

\subsection{Time Integration}
To solve the ODE \eqref{eq:semi_discrete}, explicit methods are computationally inexpensive per time step, but require very small time steps. Implicit methods generally permit larger time steps, but they require the solution of large nonlinear systems at every step, which is expensive for high-order spatial discretizations such as DFR~\citep{bidekl:26}. When implicit solvers are not efficient, such as here, exponential integrators are an attractive alternative. These have improved stability relative to explicit methods and a higher cost per step. By treating the linear dynamics exactly, exponential integrators can propagate purely linear modes without artificial damping or phase errors. For atmospheric modelling, this implies a more accurate representation of the dispersion relations of relevant wave modes, while permitting time steps based on the physical time scale of interest rather than restrictive stability limits. The present study employs the third-order exponential propagation iterative method EPI3 \citep{TOKMAN2006748}, written as 

 \begin{equation}
    \widehat{\mathbf{q}}^{\,n+1}
    =
    \widehat{\mathbf{q}}^{\,n}
    +
    \Delta t\,
    \varphi_1\!\left(\Delta t\,\mathbf{J}^{\,n}\right)
    \mathcal{F}\!\left(\widehat{\mathbf{q}}^{\,n}\right)
    +
    \frac{2}{3}\Delta t\,
    \varphi_2\!\left(\Delta t\,\mathbf{J}^{\,n}\right)
    \mathbf{R}^{\,n-1},
    \label{eq:epi3}
\end{equation}
where
\begin{equation*}
    \varphi_1(\mathbf{A})
    =
    \left(e^{\mathbf{A}}-\mathbf{I}\right)\mathbf{A}^{-1},
    \qquad
    \varphi_2(\mathbf{A})
    =
    \left(e^{\mathbf{A}}-\mathbf{I}-\mathbf{A}\right)\mathbf{A}^{-2},
    \label{eq:epi_phi_functions}
\end{equation*}
and
\begin{equation*}
    \mathbf{J}^{\,n}
    =
    \left.
    \frac{\partial \mathcal{F}}
    {\partial \widehat{\mathbf{q}}}
    \right|_{\widehat{\mathbf{q}}^{\,n}},
    \qquad
    \mathbf{R}^{\,n-1}
    =
    \mathcal{F}\!\left(\widehat{\mathbf{q}}^{\,n-1}\right)
    -
    \mathcal{F}\!\left(\widehat{\mathbf{q}}^{\,n}\right)
    -
    \mathbf{J}^{\,n}
    \left(
        \widehat{\mathbf{q}}^{\,n-1}
        -
        \widehat{\mathbf{q}}^{\,n}
    \right).
    \label{eq:epi_remainder}
\end{equation*}

The actions of $\varphi_1\!\left(\Delta t\,\mathbf{J}^{\,n}\right)$ and
$\varphi_2\!\left(\Delta t\,\mathbf{J}^{\,n}\right)$ on the vectors
$\mathcal{F}\!\left(\widehat{\mathbf{q}}^{\,n}\right)$ and
$\mathbf{R}^{\,n-1}$ can be simultaneously computed in a single Krylov
subspace projection using the methodology presented in
\citep{GAUDREAULT2018236}. The Jacobian-vector products required to
construct the Krylov subspace are approximated using finite differences. For all simulations, the Krylov subspace approximation used in the exponential time integrator employs a tolerance of $10^{-12}$ and a subspace dimension of 10.
Further implementation details can be found in
\citep{gaudreault2022high, gaudreault2026application}.

\subsection{Numerical Flux Functions}\label{flux}
The numerical flux plays a key role in the spatial discretization by defining a common flux at element interfaces from the left and right extrapolated states, denoted by $\mathbf{q}_L$ and $\mathbf{q}_R$, respectively. In this study, the Rusanov flux \citep{RUSANOV1962304} is used as a representative of a conventional upwind flux and is compared with the low-Mach-number AUSM$^+$-up formulation \citep{LIOU2006137}.

\subsubsection{Rusanov Flux}
For a one-dimensional interface normal to the $x$-direction, the Rusanov flux is written as
\begin{equation}
    \mathbf{f}^{*}_{\mathrm{Rus}}
    =
    \frac{1}{2}
    \left[
        \mathbf{F}_x(\mathbf{q}_{L})
        +
        \mathbf{F}_x(\mathbf{q}_{R})
    \right]
    -
    \frac{\lambda_{\max}}{2}
    \left(
        \mathbf{q}_{R}-\mathbf{q}_{L}
    \right),
    \label{eq:rusanov}
\end{equation}
where
\begin{equation*}
    \lambda_{\max}
    =
    \max\left(
        |u_L|+c_L,\,
        |u_R|+c_R
    \right)
\end{equation*}
is an estimate of the largest characteristic speed at the interface, and $c$ denotes the speed of sound.

The Rusanov flux is robust but becomes excessively dissipative in the low-Mach-number regime. Its dissipative contribution is proportional to the largest characteristic speed, which for the compressible Euler equations scales as $|u|+c$. For a flow with characteristic advective velocity $U$ and Mach number $M=U/c$, the ratio of the acoustic to advective velocity scales is
\begin{equation*}
    \frac{c}{U}=\frac{1}{M}.
\end{equation*}
Hence, as $M\rightarrow0$, the dissipation introduced by the Rusanov flux is governed by a velocity scale that is increasingly larger than the characteristic velocity of the physical motion. In nondimensional variables scaled by the advective velocity $U$, this corresponds to an acoustic characteristic speed of $O(M^{-1})$. Consequently, the numerical dissipation can dominate the slowly evolving real dynamics as the Mach number decreases.

This mismatch also affects the pressure scaling. For low-Mach-number flows, the pressure admits an asymptotic expansion of the form
\begin{equation}
    p
    =
    p^{(0)}
    +
    M^2 p^{(2)}
    +
    O(M^4),
    \label{eq:low_mach_pressure_scaling}
\end{equation}
where $p^{(0)}$ is spatially uniform at leading order, and the dynamically relevant pressure fluctuations are therefore of order $M^2$. Standard compressible upwind discretizations do not, in general, preserve this asymptotic scaling. Because their numerical dissipation is proportional to the acoustic wave speed, the discrete momentum balance can instead introduce pressure fluctuations of order $M$ \citep{Guillard1999OnTB}. The numerical pressure perturbation can therefore become larger than the physical $O(M^2)$ perturbation as the Mach number decreases. This is the origin of the excessive damping and loss of accuracy observed with conventional upwind fluxes in very low-Mach-number flows.

\subsubsection{AUSM$^+$-up Flux}
To obtain a numerical dissipation that remains appropriate in this regime, we employ the AUSM$^+$-up flux \citep{LIOU2006137}. The Advective Upstream Splitting Method (AUSM) family separates the interface flux into an advective contribution, transported by an interface mass flux, and a pressure contribution. For the conserved-variable formulation adopted here, the numerical flux in the $x$-direction can be written as
\begin{equation}
    \mathbf{f}^{*}_{\mathrm{AUSM}}
    =
    \dot{m}_{1/2}\,
    \boldsymbol{\psi}_{1/2}
    +
    \mathbf{p}_{1/2},
    \label{eq:ausm_flux}
\end{equation}
where
\begin{equation}
    \boldsymbol{\psi}
    =
    \begin{pmatrix}
        1\\
        u\\
        v\\
        w\\
        \theta
    \end{pmatrix},
    \qquad
    \mathbf{p}_{1/2}
    =
    \begin{pmatrix}
        0\\
        p_{1/2}\\
        0\\
        0\\
        0
    \end{pmatrix}.
\end{equation}
Here, $\dot{m}_{1/2}$ is the interface mass flux and $p_{1/2}$ is the interface pressure. The advected state $\boldsymbol{\psi}_{1/2}$ is determined by the sign of the interface mass flux,
\begin{equation}
    \boldsymbol{\psi}_{1/2}
    =
    \begin{cases}
        \boldsymbol{\psi}_{L},
        & \dot{m}_{1/2} > 0, \\[3pt]
        \boldsymbol{\psi}_{R},
        & \dot{m}_{1/2} \leq 0.
    \end{cases}
\end{equation}

The mass flux is constructed from an interface Mach number $M_{1/2}$ according to
\begin{equation}
    \dot{m}_{1/2}
    =
    c_{1/2} M_{1/2}
    \begin{cases}
        \rho_L,
        & M_{1/2} > 0, \\[3pt]
        \rho_R,
        & M_{1/2} \leq 0,
    \end{cases}
    \label{eq:ausm_mass_flux}
\end{equation}
where $c_{1/2}$ is the interface speed of sound. The left and right Mach numbers are defined using the normal velocities,
\begin{equation}
    M_L=\frac{u_L}{c_{1/2}},
    \qquad
    M_R=\frac{u_R}{c_{1/2}}.
\end{equation}

In AUSM$^+$-up, the interface speed of sound $c_{1/2}$ can be defined using the critical acoustic speed as $c_{1/2}=\min(\bar{c}_L,\bar{c}_R)$, where $\bar{c}=c^{*2}/\max(c^*,|u|)$ and $c^*$ denotes the critical speed of sound evaluated at a local Mach number of unity. Alternatively, the interface speed of sound can be determined using an appropriate numerical average. The latter approach is adopted in this study. The interface Mach number is given by

\begin{equation}
    M_{1/2}
    =
    \mathcal{M}^{+}_{(4)}(M_L)
    +
    \mathcal{M}^{-}_{(4)}(M_R)
    -
    \frac{K_p}{f_a}
    \max\left(1-\sigma\bar{M}^{2},0\right)
    \frac{p_R-p_L}
    {\rho_{1/2}c_{1/2}^{2}},
    \label{eq:ausm_interface_mach}
\end{equation}
with
\begin{equation}
    \rho_{1/2}
    =
    \frac{\rho_L+\rho_R}{2},
    \qquad
    \bar{M}^{2}
    =
    \frac{u_L^2+u_R^2}{2c_{1/2}^{2}}.
    \label{eq:ausm_average_mach}
\end{equation}
The final term in \eqref{eq:ausm_interface_mach} is a pressure-diffusion term, commonly referred to as the $p$-term. It introduces pressure-velocity coupling into the mass flux and plays an important role in the low-Mach-number behaviour of the scheme.

The interface pressure is defined as
\begin{equation}
    p_{1/2}
    =
    \mathcal{P}^{+}_{(5)}(M_L)p_L
    +
    \mathcal{P}^{-}_{(5)}(M_R)p_R
    -
    K_u
    \mathcal{P}^{+}_{(5)}(M_L)
    \mathcal{P}^{-}_{(5)}(M_R)
    f_a c_{1/2}
    (\rho_L+\rho_R)
    (u_R-u_L).
    \label{eq:ausm_pressure_flux}
\end{equation}
The final contribution in \eqref{eq:ausm_pressure_flux} is the velocity-diffusion, or $u$-term. Its multiplication by the Mach-number-dependent factor $f_a$ is particularly important at low Mach number: as $M\rightarrow0$, the velocity-dependent pressure dissipation is reduced together with the characteristic advective scale instead of remaining proportional to the acoustic speed.

The scaling function is defined as
\begin{equation}
    f_a=M_0(2-M_0),
    \label{eq:fa_scaling}
\end{equation}
where
\begin{equation}
    M_0
    =
    \min\left[
        1,\,
        \max\left(
            \bar{M},
            M_{\mathrm{cut}}
        \right)
    \right].
    \label{eq:M0_scaling}
\end{equation}
The cut-off Mach number $M_{\mathrm{cut}}$ prevents the scaling factor from vanishing and limits the magnitude of the low-Mach correction. In this study, $M_{\mathrm{cut}}=10^{-13}$ is used for the velocity-diffusion scaling. The treatment of the pressure-diffusion scaling is discussed below.

The functions $\mathcal{M}^{\pm}_{(4)}$ and $\mathcal{P}^{\pm}_{(5)}$ are the fourth- and fifth-order Mach-number and pressure-splitting polynomials of AUSM$^+$, respectively. The complete piecewise definitions, including the supersonic branches, can be found in \citep{LIOU2006137}. Since the atmospheric flows considered in this work are subsonic, $|M_L|<1$ and $|M_R|<1$, only their subsonic branches are required. They may therefore be written as
\begin{equation}
    \mathcal{M}_{(2)}^{\pm}(M)
    =
    \pm\frac{1}{4}(M\pm1)^2,
    \label{eq:ausm_M2}
\end{equation}
\begin{equation}
    \mathcal{M}_{(4)}^{\pm}(M)
    =
    \mathcal{M}_{(2)}^{\pm}(M)
    \left[
        1
        \mp
        16\beta
        \mathcal{M}_{(2)}^{\mp}(M)
    \right],
    \qquad
    \beta=\frac{1}{8},
    \label{eq:ausm_M4}
\end{equation}
and
\begin{equation}
    \mathcal{P}_{(5)}^{\pm}(M)
    =
    \mathcal{M}_{(2)}^{\pm}(M)
    \left[
        (\pm2-M)
        \mp
        16\alpha M
        \mathcal{M}_{(2)}^{\mp}(M)
    \right],
    \qquad
    \alpha=\frac{3}{16}.
    \label{eq:ausm_P5}
\end{equation}
For the parameters, the standard values are used: $K_p = 0.25, \quad K_u = 0.75, \quad \sigma = 1.0$.

The AUSM$^+$-up construction can be interpreted as modifying the two components of the pressure-velocity coupling separately. The $p$-term in \eqref{eq:ausm_interface_mach} controls the pressure contribution to the mass flux, whereas the $u$-term in \eqref{eq:ausm_pressure_flux} controls velocity-dependent dissipation in the interface pressure. Their Mach-number scaling allows the numerical dissipation to decrease appropriately as the incompressible limit is approached, while retaining the pressure-velocity coupling required for a robust compressible formulation.

For the nearly hydrostatic atmospheric flows considered in the present study, a small modification is applied to AUSM$^+$-up following \citet{Miczek2013}. The factor $f_a$ in the $p$-term of \eqref{eq:ausm_interface_mach} is set to unity, while the Mach-number-dependent $f_a$ scaling is retained in the $u$-term of \eqref{eq:ausm_pressure_flux}. Denoting these factors separately by $f_a^p$ and $f_a^u$, this corresponds to $f_a^p=1$ and $f_a^u=f_a$. Setting $f_a^p=1$ removes the cutoff dependence from $p$-term, as demonstrated by the hydrostatic flux tests of \citet{Miczek2013}. Throughout the remainder of this paper, this modified AUSM$^+$-up flux is referred to as AUSM$^+$-upH.

\section{Deviation Method} \label{DM}

Atmospheric flow models typically include external effects such as gravity, rotation, and geometric forcing. These effects generally enter the governing equations as source terms, thereby converting the conservation law into a balance law. For a hydrostatic state, the flux-gradient term is exactly balanced by the gravity source term at the continuous level, but their discrete approximations may not cancel exactly. In the DFR formulation, flux divergence relies on polynomial derivatives with interface corrections, while source terms are evaluated locally at solution points. Consequently, a continuous balance between pressure-gradient and gravitational terms does not guarantee exact discrete cancellation. This imbalance can generate spurious motion. For nearly hydrostatic flows, which may be viewed as small perturbations about a hydrostatic reference state, the true imbalance between the flux-gradient and source terms is very small. Therefore, it can be easily contaminated by discretization errors.

Numerical techniques designed to address this issue are commonly referred to as well-balanced methods. In this section, we briefly describe a simple deviation-based well-balanced formulation following \citet{Berberich_Chandrashekar_Klingenberg_2021}, similar to the one from \cite{dedner:04}. The underlying idea of evolving perturbations relative to a prescribed background state has a much longer history in atmospheric modelling \citep{Clark1977}. This method is adopted here because of its generality for both stationary and time-dependent flows, as well as its relative ease of implementation.

The deviation method is built around a prescribed reference solution, denoted by $\tilde{\mathbf{q}}$, which is assumed to be known in advance. In the usual well-balanced setting, this reference state is a stationary solution of Eq.~\eqref{eq:pde}. Hence, it satisfies the steady balance relation
\begin{equation}
    \frac{\partial \mathbf{F}_x(\widetilde{\mathbf{q}})}{\partial x}
    +
    \frac{\partial \mathbf{F}_y(\widetilde{\mathbf{q}})}{\partial y}
    +
    \frac{\partial \mathbf{F}_z(\widetilde{\mathbf{q}})}{\partial z}
    =
    \mathbf{S}(\widetilde{\mathbf{q}}).
    \label{eq:steady_balance_reference}
\end{equation}
However, the deviation method can also be formulated for a known time-dependent target solution. Throughout this discussion, we refer to $\tilde{\mathbf{q}}$ as the target solution.

The main idea is to evolve the difference between the numerical solution and the prescribed target state, rather than the full solution itself. To derive the corresponding equation, we first evaluate Eq.~\eqref{eq:pde} at $\tilde{\mathbf{q}}$ and subtract it from the original balance law. This gives
\begin{equation}
\begin{aligned}
    &
    \left(
    \frac{\partial \mathbf{q}}{\partial t}
    +
    \frac{\partial \mathbf{F}_x(\mathbf{q})}{\partial x}
    +
    \frac{\partial \mathbf{F}_y(\mathbf{q})}{\partial y}
    +
    \frac{\partial \mathbf{F}_z(\mathbf{q})}{\partial z}
    -
    \mathbf{S}(\mathbf{q})
    \right)
    \\
    &\quad
    -
    \left(
    \frac{\partial \widetilde{\mathbf{q}}}{\partial t}
    +
    \frac{\partial \mathbf{F}_x(\widetilde{\mathbf{q}})}{\partial x}
    +
    \frac{\partial \mathbf{F}_y(\widetilde{\mathbf{q}})}{\partial y}
    +
    \frac{\partial \mathbf{F}_z(\widetilde{\mathbf{q}})}{\partial z}
    -
    \mathbf{S}(\widetilde{\mathbf{q}})
    \right)
    =
    0.
\end{aligned}
\label{eq:deviation_subtraction}
\end{equation}

After rearranging terms, one obtains
\begin{equation}
\frac{\partial}{\partial t}\left(\mathbf{q}-\widetilde{\mathbf{q}}\right)
+
\frac{\partial}{\partial x}\left[\mathbf{F}_x(\mathbf{q})-\mathbf{F}_x(\widetilde{\mathbf{q}})\right]
+
\frac{\partial}{\partial y}\left[\mathbf{F}_y(\mathbf{q})-\mathbf{F}_y(\widetilde{\mathbf{q}})\right]
+
\frac{\partial}{\partial z}\left[\mathbf{F}_z(\mathbf{q})-\mathbf{F}_z(\widetilde{\mathbf{q}})\right]
=
\mathbf{S}(\mathbf{q})-\mathbf{S}(\widetilde{\mathbf{q}}).
\label{eq:deviation_balance_full}
\end{equation}
Introducing the deviation variable
\begin{equation*}
    \Delta \mathbf{q} := \mathbf{q}-\widetilde{\mathbf{q}},
    \label{eq:deviation_variable}
\end{equation*}
or equivalently $\mathbf{q}=\Delta \mathbf{q}+\tilde{\mathbf{q}}$, the balance law becomes
\begin{equation}
\begin{aligned}
\frac{\partial \Delta \mathbf{q}}{\partial t}
&+
\frac{\partial}{\partial x}
\left[
\mathbf{F}_x(\Delta \mathbf{q}+\widetilde{\mathbf{q}})
-
\mathbf{F}_x(\widetilde{\mathbf{q}})
\right]
+
\frac{\partial}{\partial y}
\left[
\mathbf{F}_y(\Delta \mathbf{q}+\widetilde{\mathbf{q}})
-
\mathbf{F}_y(\widetilde{\mathbf{q}})
\right]
+
\frac{\partial}{\partial z}
\left[
\mathbf{F}_z(\Delta \mathbf{q}+\widetilde{\mathbf{q}})
-
\mathbf{F}_z(\widetilde{\mathbf{q}})
\right]
\\
&=
\mathbf{S}(\Delta \mathbf{q}+\widetilde{\mathbf{q}})
-
\mathbf{S}(\widetilde{\mathbf{q}}).
\end{aligned}
\label{eq:deviation_method_full}
\end{equation}

Thus, the numerical method is formulated in terms of the deviation from the target state. If the computed solution coincides with the target solution, then $\Delta \mathbf{q}=0$, and both the flux difference and source difference vanish. Consequently, the reference state is preserved exactly. For small perturbations about this state, the method removes the large background balance from the residual, allowing the numerical evolution to be driven primarily by the perturbation dynamics.

\section{Numerical Experiments} 
\label{results}

The numerical experiments consider two computational geometries. The Gresho vortex test (Section~\ref{sec:gresho}), the isothermal hydrostatic-equilibrium test (Section~\ref{sec:hydrostatic_test}) and the rising-thermal-bubble test (Section~\ref{sec:thermal_bubble}) are conducted in a two-dimensional Cartesian domain, whereas the steady-state atmosphere over orography (Section~\ref{sec:orography}) and the non-hydrostatic gravity-wave test (Section~\ref{sec:gravitywave}) are performed as three-dimensional global simulations on a cubed-sphere grid \citep{Sadourny1972}.

For the two-dimensional Cartesian experiments, the governing equations presented in Section~\ref{equations} are reduced to the $x$-$z$ plane. In the three-dimensional global simulations, the same compressible Euler system is formulated in curvilinear coordinates on the cubed-sphere grid, with potential temperature retained as the prognostic thermodynamic variable. A complete description of the Euler equations in curvilinear coordinates, together with the corresponding implementation details, is provided in~\citep{gaudreault2026application}.

For the cubed-sphere experiments, the horizontal resolution $\Delta_h$ is
estimated following~\citet{gaudreault2022high} as
\begin{equation*}
    \Delta_h \simeq \frac{90^\circ}{N_e N_s},
\end{equation*}
where $N_e$ denotes the number of elements along each horizontal direction
of a cubed-sphere panel and $N_s$ is the number of solution points per
element. For a DFR polynomial of degree $p$, $N_s=p+1$.

At the upper and lower boundaries of the three-dimensional domain, the contravariant vertical velocity satisfies the impermeability condition $u^3=0$. For experiments involving topography, the lower coordinate surface conforms to the prescribed terrain.

\subsection{Gresho Vortex} \label{sec:gresho}

The Gresho vortex test case \citep{liska2003comparison, Miczek2013} is used to assess the performance of the aforementioned flux functions under low Mach-number conditions and without gravity. A more detailed theoretical discussion on the low-Mach number behaviour can be found in \citep{Jung2024}.

This test involves a stationary rotating vortex, where the centrifugal forces are exactly balanced by pressure gradients. The vortex is centred at $(x_c,z_c)=(0.5,0.5)$ in the unit square, with periodic boundary conditions imposed in both spatial directions. The domain is filled with an ideal gas of constant density ($\rho=1$). The rotational motion of the vortex is prescribed by the following tangential velocity profile:

\begin{equation}
u_\phi(r)=
\begin{cases}
5r, & r<0.2,\\
2-5r, & 0.2\le r<0.4,\\
0, & r\ge 0.4,
\end{cases}
\end{equation}
where the radial distance from the vortex centre is defined as

\begin{equation*}
    r=\sqrt{(x-x_c)^2+(z-z_c)^2}.
\end{equation*}
The tangential velocity reaches its maximum value, $u_{\phi,\max}=1$, at $r=0.2$.

The pressure profile is chosen such that the resulting pressure gradient provides the centripetal force required to maintain the rotating vortex:

\begin{equation}
p(r)=
\begin{cases}
p_{\mathrm{base}}+12.5r^2, & r<0.2,\\
p_{\mathrm{base}}+4\ln(5r)+4-20r+12.5r^2, & 0.2\le r<0.4,\\
p_{\mathrm{base}}+4\ln(2)-2, & r\ge 0.4.
\end{cases}
\end{equation}

The pressure profile is continuous and differentiable throughout the domain. The background pressure constant $p_{\mathrm{base}}$ is set to

\begin{equation*}
    p_{\mathrm{base}}
    =
    \frac{1}{\gamma M_{\max}^2}
    -
    \frac{1}{2},
    \qquad
    \gamma=1.4.
\end{equation*}

Here, $M_{\max}$ denotes the maximum initial Mach number, which occurs at $r=0.2$. The temperature is initialized from the ideal-gas equation of state, and the initial conditions are specified using dimensional quantities.

By varying $M_{\max}$, the background pressure is adjusted while the prescribed velocity profile remains unchanged. This makes the test suitable for assessing the ability of a numerical method to preserve the vortex structure as the Mach number decreases.

\begin{figure}[htbp]
    \centering

    \begin{subfigure}{0.32\textwidth}
        \centering
        \includegraphics[width=\textwidth]{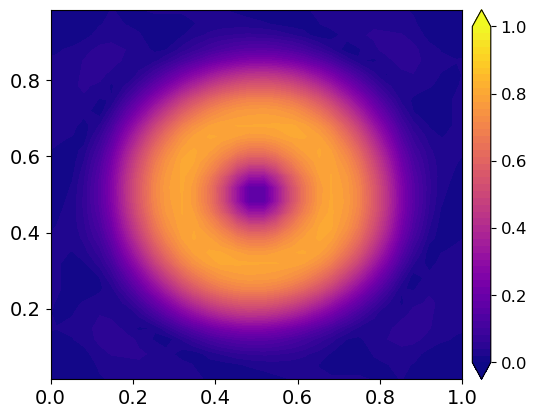}
        \caption{$M_{max}=10^{-1}$}
    \end{subfigure}
    \hfill
    \begin{subfigure}{0.32\textwidth}
        \centering
        \includegraphics[width=\textwidth]{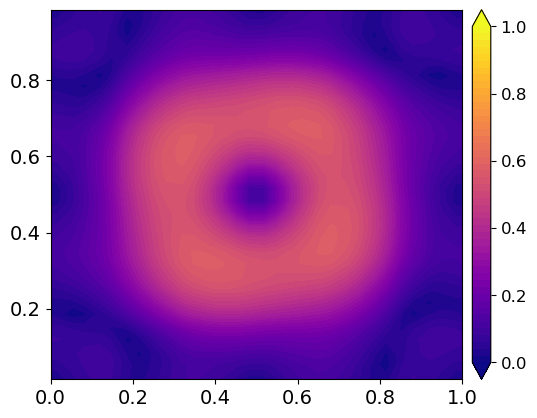}
        \caption{$M_{max}=10^{-2}$}
    \end{subfigure}
    \hfill
    \begin{subfigure}{0.32\textwidth}
        \centering
        \includegraphics[width=\textwidth]{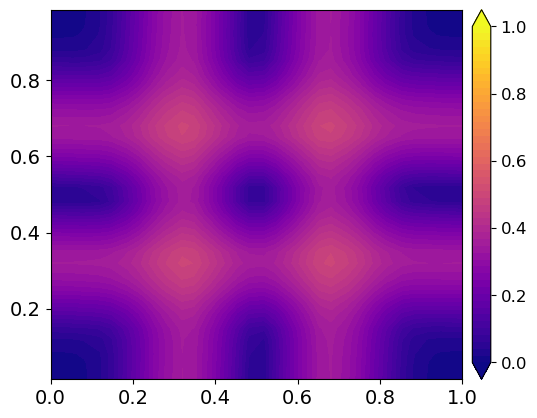}
        \caption{$M_{max}=10^{-3}$}
    \end{subfigure}

    \caption{Numerical solution of Gresho vortex after one full revolution ($t \approx 1.26$) with Rusanov flux function. The magnitude of the velocity $|V|$ for different Mach numbers is shown. A $15 \times 15$ grid is used with polynomial order $p=1$ and $\Delta t = 0.01$.}
    \label{fig:mavar}
\end{figure}

Figure~\ref{fig:mavar} illustrates the results obtained with the Rusanov flux function for different Mach numbers. As expected, the solution becomes increasingly diffusive as the Mach number decreases, eventually developing a cross-shaped pattern for $M_{\max}=10^{-3}$. This issue occurs because the flux dissipation is controlled by the fastest wave in the system, and at low Mach numbers this wave speed is dominated by the speed of sound. Although the vortex velocity remains fixed, reducing the Mach number increases the sound speed, leading to excessive numerical diffusion in the solution.

\begin{figure}[htbp]
    \centering

    \begin{subfigure}{0.32\textwidth}
        \centering
        \includegraphics[width=\textwidth]{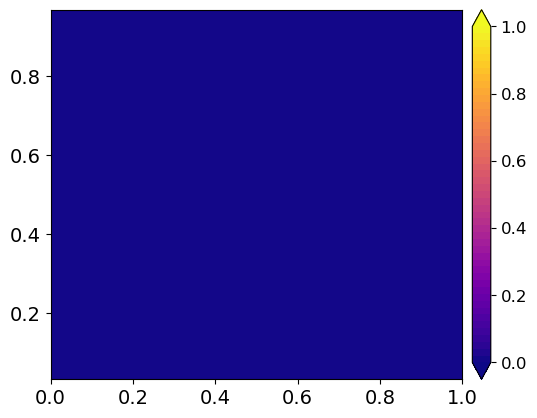}
        \caption{$p=0$}
    \end{subfigure}
    \hfill
    \begin{subfigure}{0.32\textwidth}
        \centering
        \includegraphics[width=\textwidth]{M1e-3_k_1.png}
        \caption{$p=1$}
    \end{subfigure}
    \hfill
    \begin{subfigure}{0.32\textwidth}
        \centering
        \includegraphics[width=\textwidth]{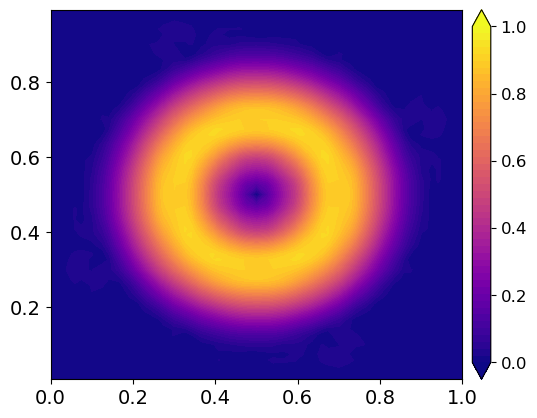}
        \caption{$p=2$}
    \end{subfigure}

    \caption{Velocity magnitude $|V|$ of the Gresho vortex after one full revolution ($t \approx 1.26$) at $M_{max} = 10^{-3}$, obtained using Rusanov flux function on a fixed $15 \times 15$ grid and $\Delta t = 0.01$ for varying polynomial degrees $p$.}
    \label{fig:pvar}
\end{figure}

As shown in Figure~\ref{fig:pvar}, the polynomial order $p=0$ (equivalent to a first-order finite-volume discretization) introduces excessive numerical diffusion which damps the solution down to near machine zero. At $p=1$, it exhibits the characteristic spurious features expected from schemes that are not low Mach-number compliant. However, the scheme performs notably better for $p=2$ than one might expect, a feature which is not unique to the present study \citep{Jung2024, barsukow-2025}. This robustness is limited, as further decreasing the Mach number once again introduces spurious structures for $p=2$.

\begin{figure}[htbp]
    \centering

    \begin{subfigure}{0.32\textwidth}
        \centering
        \includegraphics[width=\textwidth]{M1e-3_k_1.png}
        \caption{$15 \times 15$}
    \end{subfigure}
    \hfill
    \begin{subfigure}{0.32\textwidth}
        \centering
        \includegraphics[width=\textwidth]{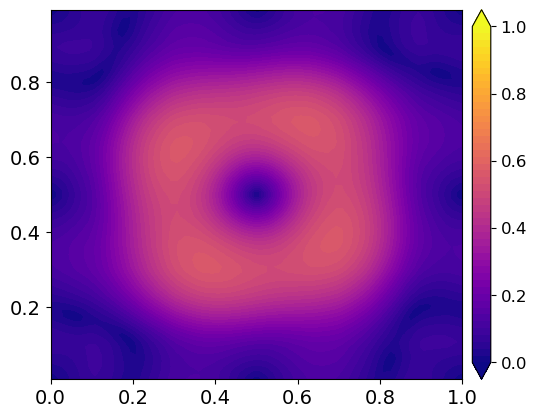}
        \caption{$30 \times 30$}
    \end{subfigure}
    \hfill
    \begin{subfigure}{0.32\textwidth}
        \centering
        \includegraphics[width=\textwidth]{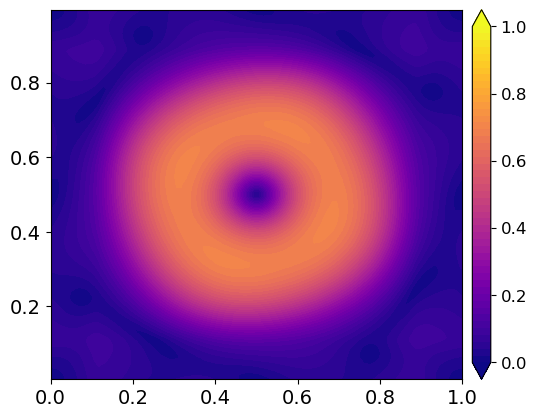}
        \caption{$45 \times 45$}
    \end{subfigure}

    \caption{
Velocity magnitude $|\mathbf{V}|$ of the Gresho vortex after one full revolution ($t \approx 1.26$) at $M_{\max}=10^{-3}$, obtained using the Rusanov flux function with polynomial degree
$p=1$. Results are shown for increasing grid resolutions and $\Delta t=0.01$.
}
    \label{fig:rvar}
\end{figure}

The results improve with increasing grid resolution, as demonstrated in Figure~\ref{fig:rvar}. It should be noted that, in this case, the required grid resolution is dictated by the Mach number of the problem, whereas spatial resolution is generally selected based on the smallest spatial scale that needs to be resolved.

\begin{figure}[htbp]
    \centering

    \begin{subfigure}{0.32\textwidth}
        \centering
        \includegraphics[width=\textwidth]{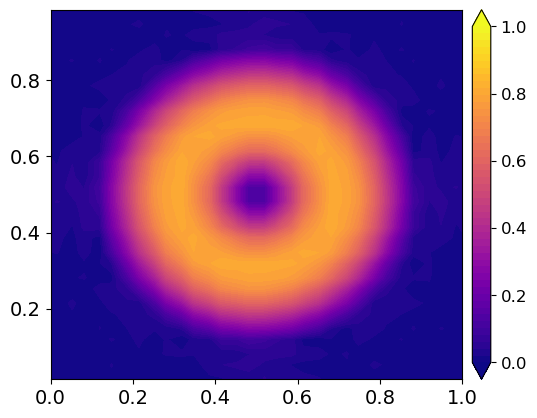}
        \caption{$M_{max}=10^{-1}$}
    \end{subfigure}
    \hfill
    \begin{subfigure}{0.32\textwidth}
        \centering
        \includegraphics[width=\textwidth]{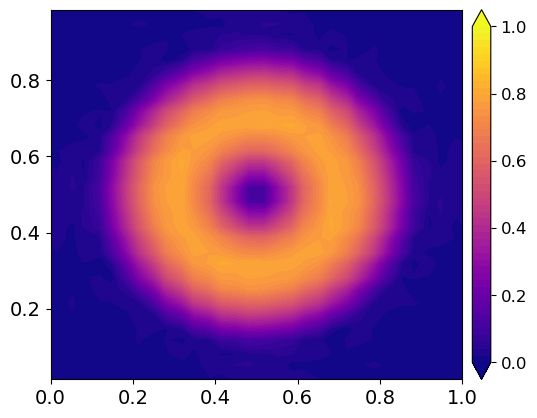}
        \caption{$M_{max}=10^{-2}$}
    \end{subfigure}
    \hfill
    \begin{subfigure}{0.32\textwidth}
        \centering
        \includegraphics[width=\textwidth]{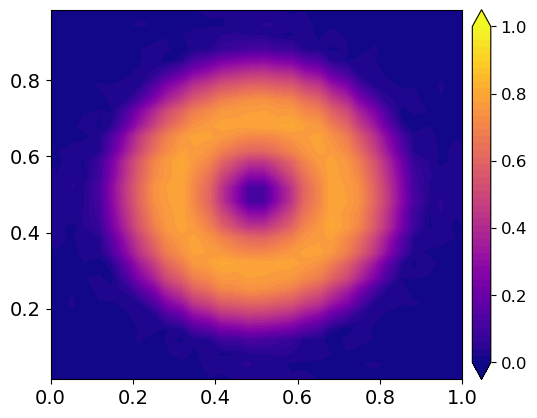}
        \caption{$M_{max}=10^{-3}$}
    \end{subfigure}

    \caption{Results obtained using the same setup as in Figure~\ref{fig:mavar}, but with the AUSM$^+$-upH flux function.}
    \label{fig:ausmMvar}
\end{figure}

The numerical experiments presented above with the Rusanov flux confirm the expected challenges associated with the low Mach-number regime. They indicate that modifications to the numerical flux function are necessary to obtain reliable solutions for low Mach-number flows. In Figure ~\ref{fig:ausmMvar}, results with AUSM$^+$-upH show good accuracy across the range of Mach numbers, indicating improved asymptotic consistency.

\subsection{Isothermal Hydrostatic Equilibrium}
\label{sec:hydrostatic_test}

This test considers a two-dimensional isothermal hydrostatic atmosphere, which should ideally remain in equilibrium without generating any spurious motion as it is advanced in time. The hydrostatic pressure profile is prescribed as
\begin{equation}
    p(z)=p_0\exp\left(-\frac{z}{H}\right),
    \qquad
    H=\frac{R_dT_0}{g}.
\end{equation}
Here, \(p_0=100{,}000~\mathrm{Pa}\), \(T_0=300~\mathrm{K}\).

The corresponding density profile is obtained from the equation of state
\begin{equation}
    \rho(z)=\frac{p(z)}{R_dT_0}
\end{equation}
and the initial potential temperature profile is computed from Equation \eqref{potential}.

These initial conditions are in exact hydrostatic balance, and the deviation formulation indeed evolves a perturbation state that is identically zero. To test the stability of perturbations to the hydrostatic balance, a perturbation at the level of machine precision is added to the horizontal velocity field.

The computational domain has a horizontal extent of \(1000~\mathrm{m}\) and a vertical extent of \(1500~\mathrm{m}\). It is discretized using a \(20\times30\) element mesh. Periodic boundary conditions are applied in the horizontal direction, while Dirichlet boundary conditions are imposed in the vertical direction. The vertical boundary values are initialized using the hydrostatic profile and are kept fixed throughout the simulation.

To assess the preservation of the hydrostatic state, we compute the maximum Mach number over all grid cells as a function of time. Since the exact equilibrium solution has zero velocity, the initially nonzero Mach number occurs due to the above-mentioned perturbation. Any subsequent growth indicates numerical errors.

\begin{figure}[htbp]
    \centering

    \begin{subfigure}{0.5\textwidth}
        \centering
        \includegraphics[width=\textwidth]{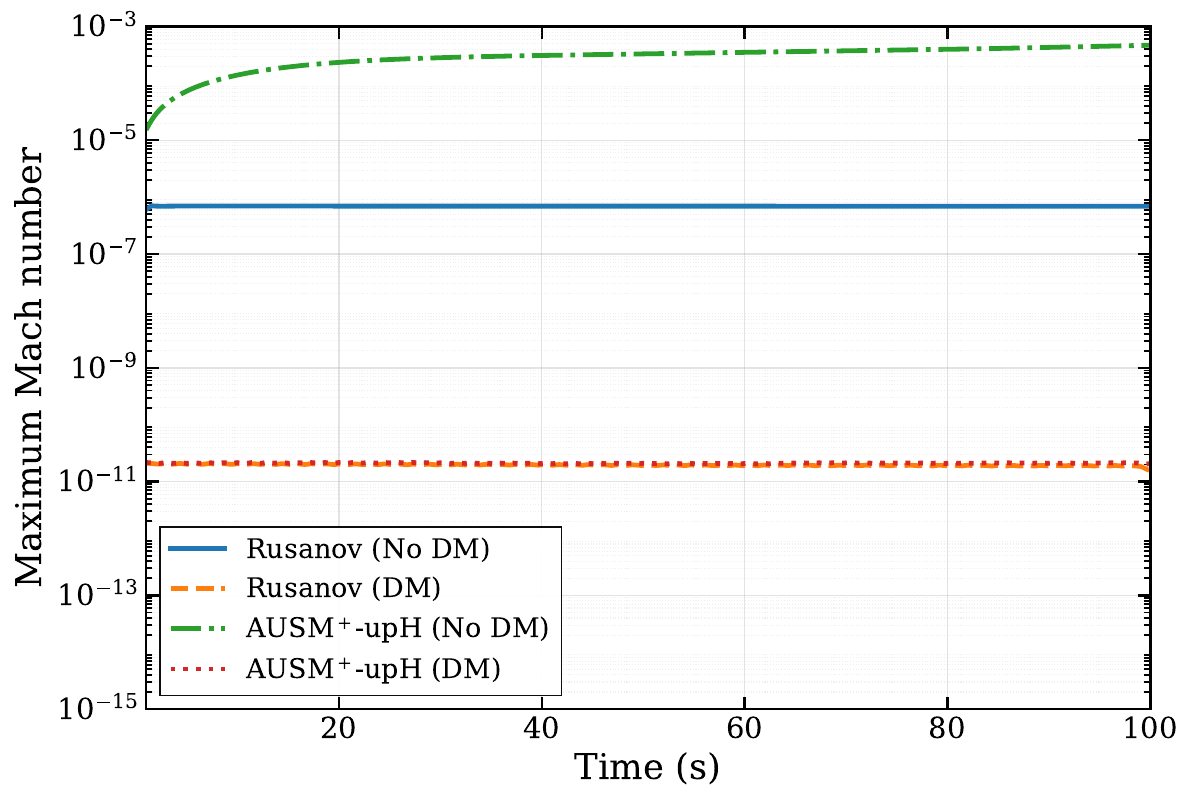}
        \caption{\(p=1\)}
    \end{subfigure}%
    \begin{subfigure}{0.5\textwidth}
        \centering
        \includegraphics[width=\textwidth]{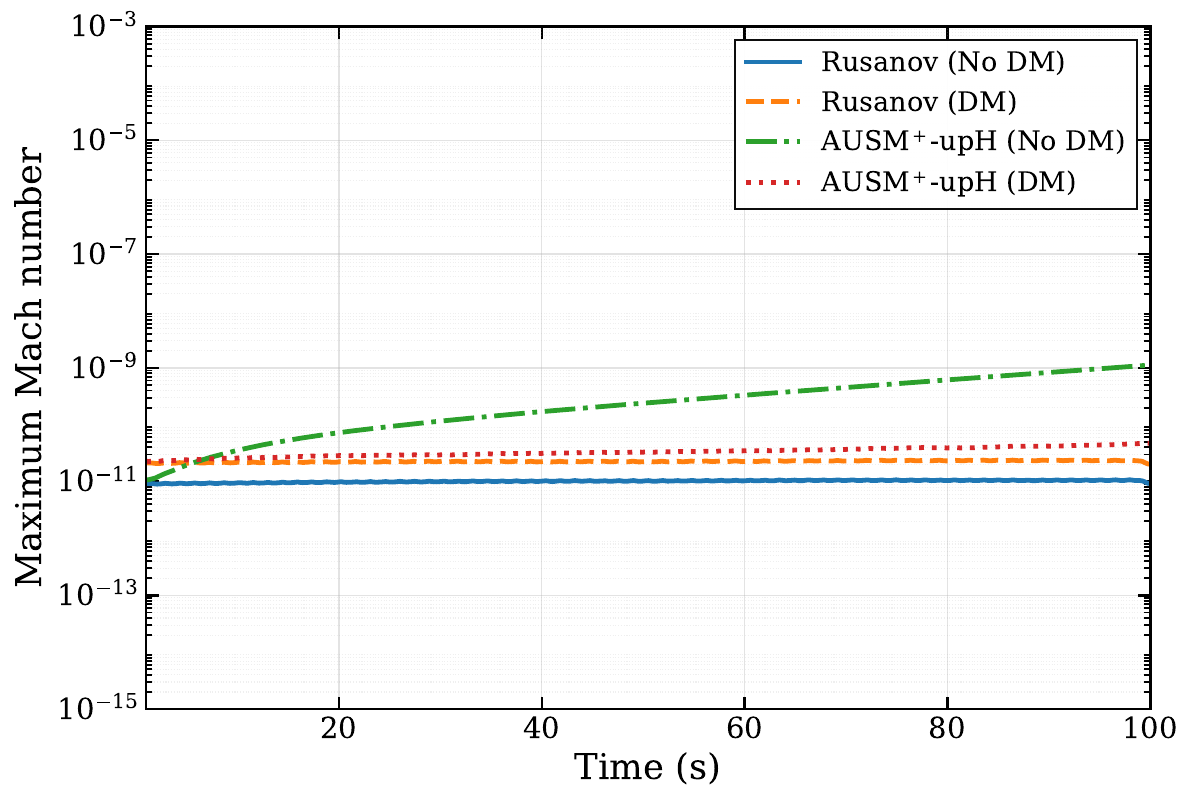}
        \caption{\(p=3\)}
    \end{subfigure}

    \caption{Evolution of the maximum Mach number in the isothermal hydrostatic equilibrium test with \(\Delta t=0.01~\mathrm{s}\) on a \(20 \times 30\) element mesh. DM denotes the deviation method. Panel (a) shows results for first-order spatial discretization, while panel (b) shows results for third-order spatial discretization. The curves in panel (b) have been smoothed for better visibility.}
    \label{fig:hyd}
\end{figure}

Figure~\ref{fig:hyd} compares the preservation of the isothermal hydrostatic equilibrium for Rusanov and AUSM$^+$-upH fluxes, with and without the deviation formulation. The results are shown for first- and third-order spatial discretizations.

For the first-order scheme, the deviation formulation substantially improves the preservation of the hydrostatic state. Both Rusanov and AUSM$^+$-upH produce much smaller spurious velocities when the equilibrium contribution is treated separately, with errors reduced by several orders of magnitude. This improvement is most apparent for the low-order discretization, where imbalance errors are otherwise more pronounced.

For the third-order scheme, all methods maintain the hydrostatic equilibrium to nearly machine precision. AUSM$^+$-upH exhibits a steady growth in the error, although its magnitude remains small.  In this case, the difference between the results obtained with and without the deviation formulation becomes negligible, indicating that the higher-order discretization already suppresses the imbalance error very effectively. Overall, AUSM$^+$-upH becomes more stable when combined with the deviation formulation.

The apparent advantage of the Rusanov flux in this stationary experiment should not be interpreted as evidence of broader robustness. In low-Mach-number flows involving active dynamics, such as convective motion or wave propagation, its strong numerical damping could suppress physically relevant features and make it less appropriate for such simulations.

\subsection{Rising Thermal Bubble}
\label{sec:thermal_bubble}
The rising thermal bubble test \citep{robert1993bubble} is used to assess the performance of the proposed schemes for buoyancy-driven atmospheric flows. This configuration is particularly relevant because the background state is initially hydrostatic, whereas the resulting motion is driven by a thermal perturbation. Since the perturbation is positive, the bubble is warmer and lighter than the surrounding atmosphere. This generates an upward buoyancy force. As a result, the initially warm perturbation rises and begins to deform as the flow develops.

\begin{figure}[htbp]
    \centering

    \begin{subfigure}{0.48\textwidth}
        \centering
        \includegraphics[width=\textwidth]{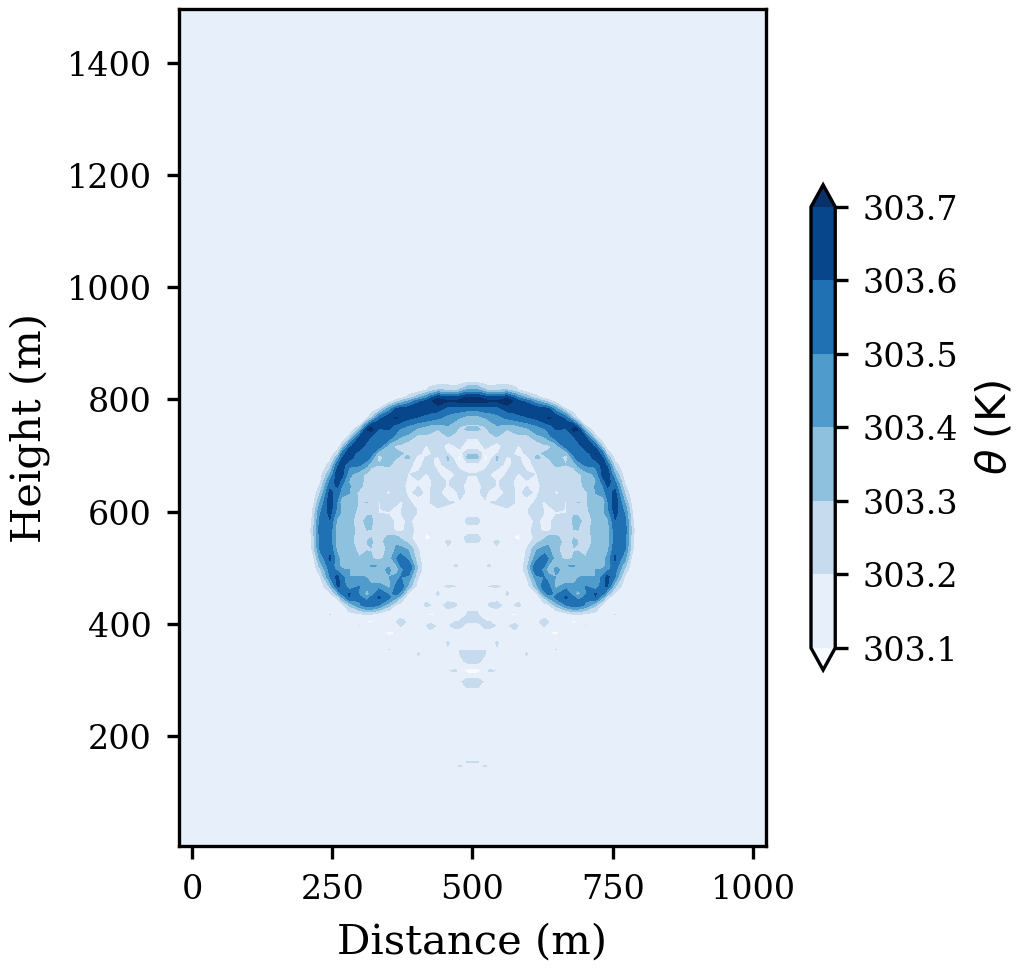}
        \caption{Rusanov (No DM)}
        \label{fig:mach_p1}
    \end{subfigure}
    \hfill
    \begin{subfigure}{0.48\textwidth}
        \centering
        \includegraphics[width=\textwidth]{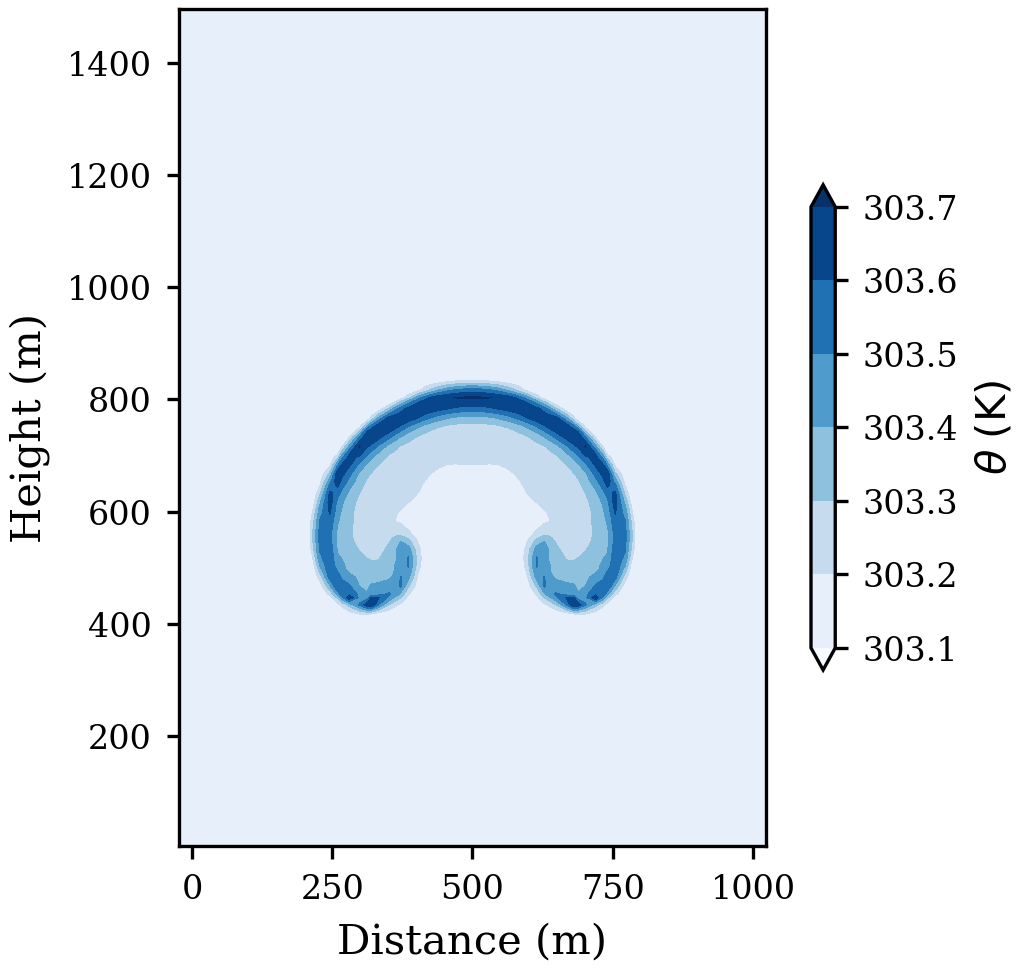}
        \caption{AUSM$^+$-upH (No DM)}
        \label{fig:mach_p2}
    \end{subfigure}

    \vspace{0.3cm}

    \begin{subfigure}{0.48\textwidth}
        \centering
        \includegraphics[width=\textwidth]{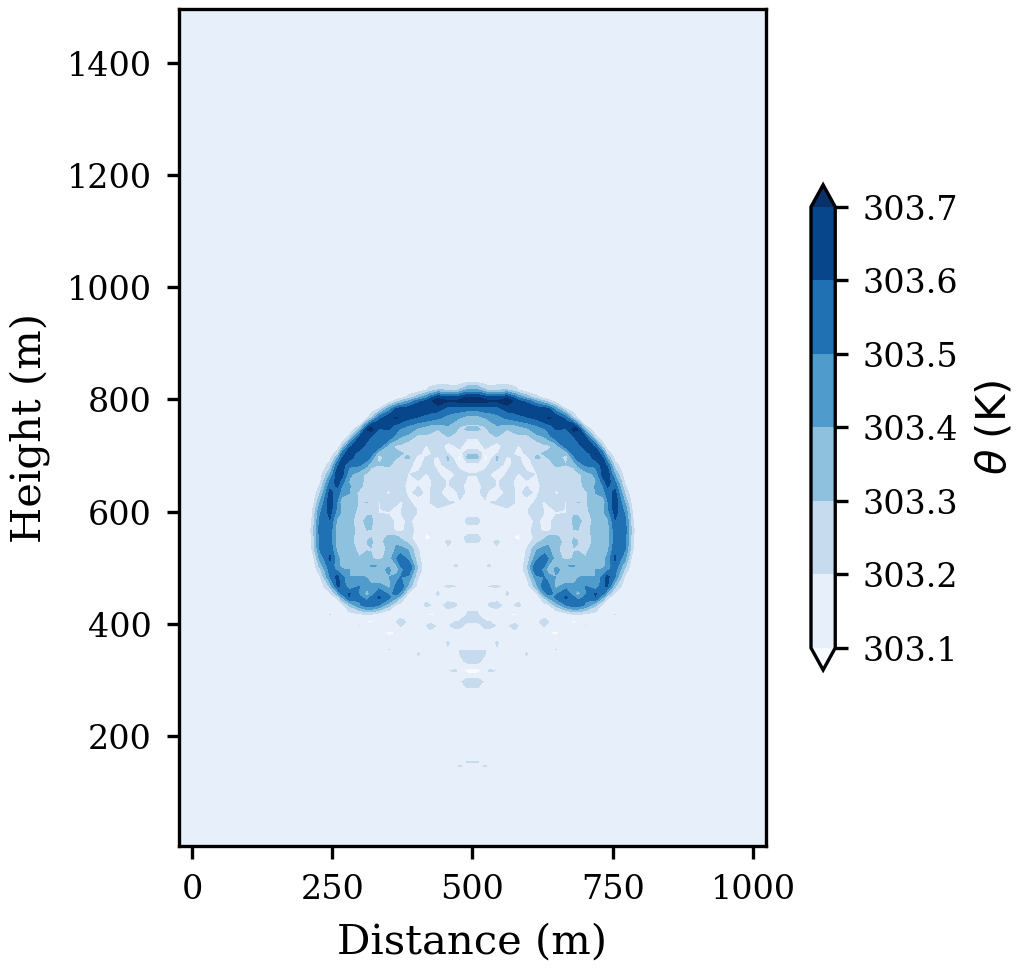}
        \caption{Rusanov (DM)}
        \label{fig:mach_p3}
    \end{subfigure}
    \hfill
    \begin{subfigure}{0.48\textwidth}
        \centering
        \includegraphics[width=\textwidth]{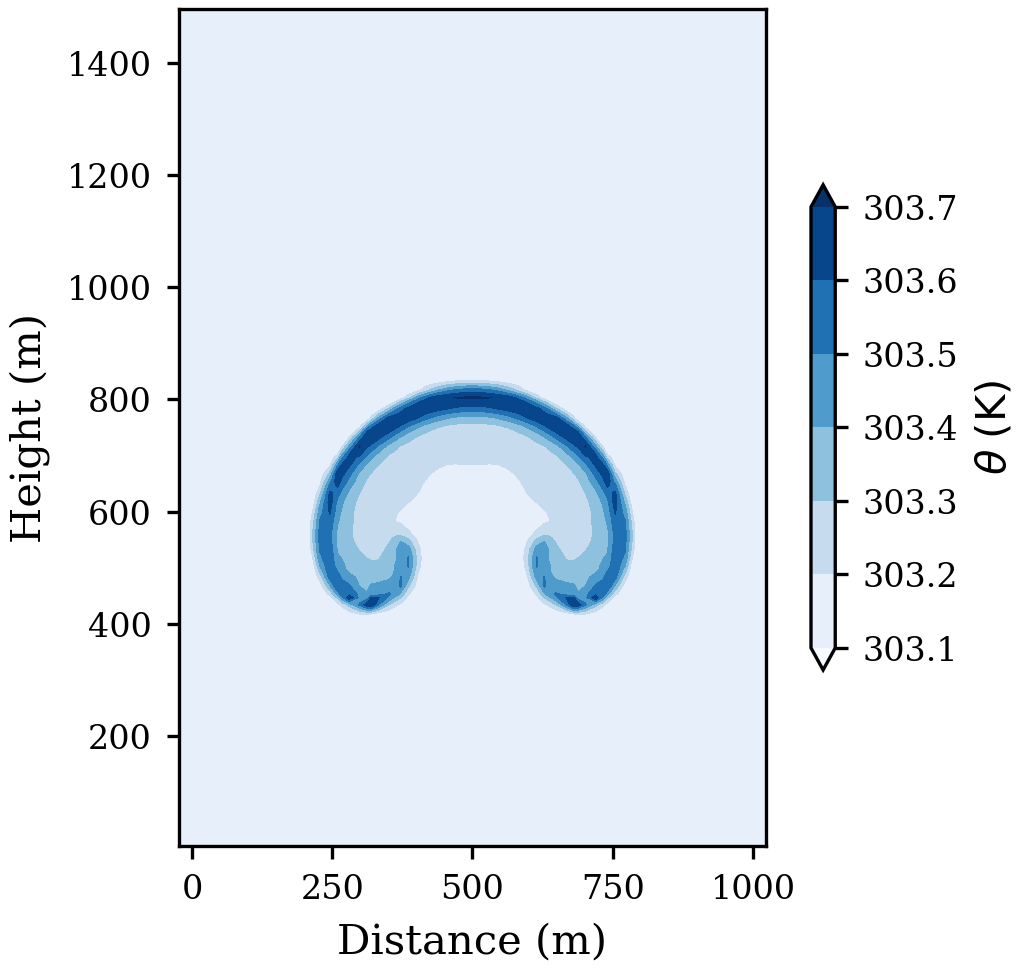}
        \caption{AUSM$^+$-upH (DM)}
        \label{fig:mach_p4}
    \end{subfigure}

    \caption{Potential-temperature contours for the rising thermal bubble at \(t=550~\mathrm{s}\). Results are shown for Rusanov and AUSM$^+$-upH fluxes with and without the deviation method (DM). The simulations use \(p=3\), a time step of \(\Delta t=5~\mathrm{s}\), and a \(19\times30\) element mesh.}
    \label{fig:bubble}
\end{figure}

The initial condition is constructed around a resting hydrostatic reference atmosphere with constant potential temperature \(\theta_0 = 303.15~\mathrm{K}\). A localized warm perturbation is then added to the potential-temperature field. The perturbation is defined as
\begin{equation}
    \theta'(x,z)
    =
    \begin{cases}
    A, & r \leq a, \\[4pt]
    A \exp\left[-\left(\dfrac{r-a}{s}\right)^2\right], & r > a,
    \end{cases}
\end{equation}
where
\begin{equation}
    r = \sqrt{(x-x_0)^2 + (z-z_0)^2}.
\end{equation}
The parameters are
\begin{equation}
    A = 0.5~\mathrm{K}, \qquad
    a = 50~\mathrm{m}, \qquad
    s = 100~\mathrm{m},
\end{equation}
and the perturbation is centred at $(x_0, z_0)= (500~\mathrm{m},260~\mathrm{m})$.
The total initial potential temperature is therefore
\begin{equation}
    \theta(x,z) = \theta_0 + \theta'(x,z).
\end{equation}

The computational domain spans \(1000~\mathrm{m}\) in the horizontal direction and \(1500~\mathrm{m}\) in the vertical direction, and is discretized using a \(19 \times 30\) element mesh. Periodic boundary conditions are used laterally, while no-flux conditions are imposed at the vertical boundaries. Time integration is performed with a time step of \(\Delta t = 5~\mathrm{s}\).

Figure~\ref{fig:bubble} presents the potential-temperature perturbation for the rising thermal bubble at \(t=550~\mathrm{s}\). The interaction between the upward motion in the bubble core and the surrounding subsidence generates shear, which deforms the initially smooth perturbation and leads to the development of a symmetric mushroom-like structure. The comparison between Rusanov and AUSM$^+$-upH shows that the low-Mach-number flux better preserves the structure of the weak buoyancy-driven flow. The Rusanov flux introduces stronger dissipation, which smooths the thermal perturbation and produces a less sharp bubble structure. However, no clear visual difference is observed between the results obtained with and without the deviation formulation in this case. This suggests that the numerical imbalance associated with the hydrostatic background state is small compared with the physical forcing generated by the warm perturbation.

\subsection{Steady-State Atmosphere over Orography}
\label{sec:orography}
The steady-state orographic test assesses the accuracy of the pressure-gradient calculation over non-flat terrain \citep{Ullrich2012DynamicalCM}. The atmosphere is initialized at rest in exact hydrostatic equilibrium. Therefore, the analytical solution remains stationary for all time, and any velocity generated during the simulation is a direct measure of numerical error.

The surface topography is prescribed as
\[
z_s(r)=
\begin{cases}
\displaystyle
\frac{h_0}{2}
\left[
1+\cos\left(\frac{\pi r}{R}\right)
\right]
\cos^2\left(\frac{\pi r}{\zeta}\right),
& r<R, \\[3mm]
0,
& r\geq R,
\end{cases}
\]
where \(h_0=2~\mathrm{km}\), \(R=3\pi/4\), and \(\zeta=\pi/16\). The quantity \(r\) is the great-circle angular distance from the mountain centre located at
\[
(\lambda_c,\phi_c)
=
\left(
\frac{3\pi}{2},0
\right).
\]

The initial velocity field is
\[
u=v=w=0.
\]
For simulations employing the deviation formulation, a machine-precision perturbation is introduced in the \(u\)-component of the horizontal velocity to avoid trivial evolution. No such perturbation is introduced in simulations without the deviation formulation.

The horizontal domain spans a full-scale Earth-sized sphere. The test uses \(N_e=30\) elements with \(N_s=3\) solution points per element in each horizontal direction, corresponding to a resolution of \(\Delta_h=1^\circ\). Vertically, 10 elements with 3 solution points per element provide 30 model levels up to a domain top of \(12\,000~\mathrm{m}\). Time integration is performed with a fixed time step of \(\Delta t=300~\mathrm{s}\).

The numerical error is quantified through the spurious velocity generated during the simulation. Since the exact solution is motionless, the maximum wind speed provides a convenient diagnostic,
\[
V_{\max}(t)
=
\max_{\Omega}
\left(
\sqrt{u^2+v^2+w^2}
\right).
\]
The maximum wind speed is evaluated over the entire computational domain at each output time, where smaller values of \(V_{\max}\) indicate a lower numerical error.

\begin{figure}[htbp]
    \centering

    \begin{subfigure}[t]{0.48\textwidth}
        \centering
        \includegraphics[width=\textwidth]{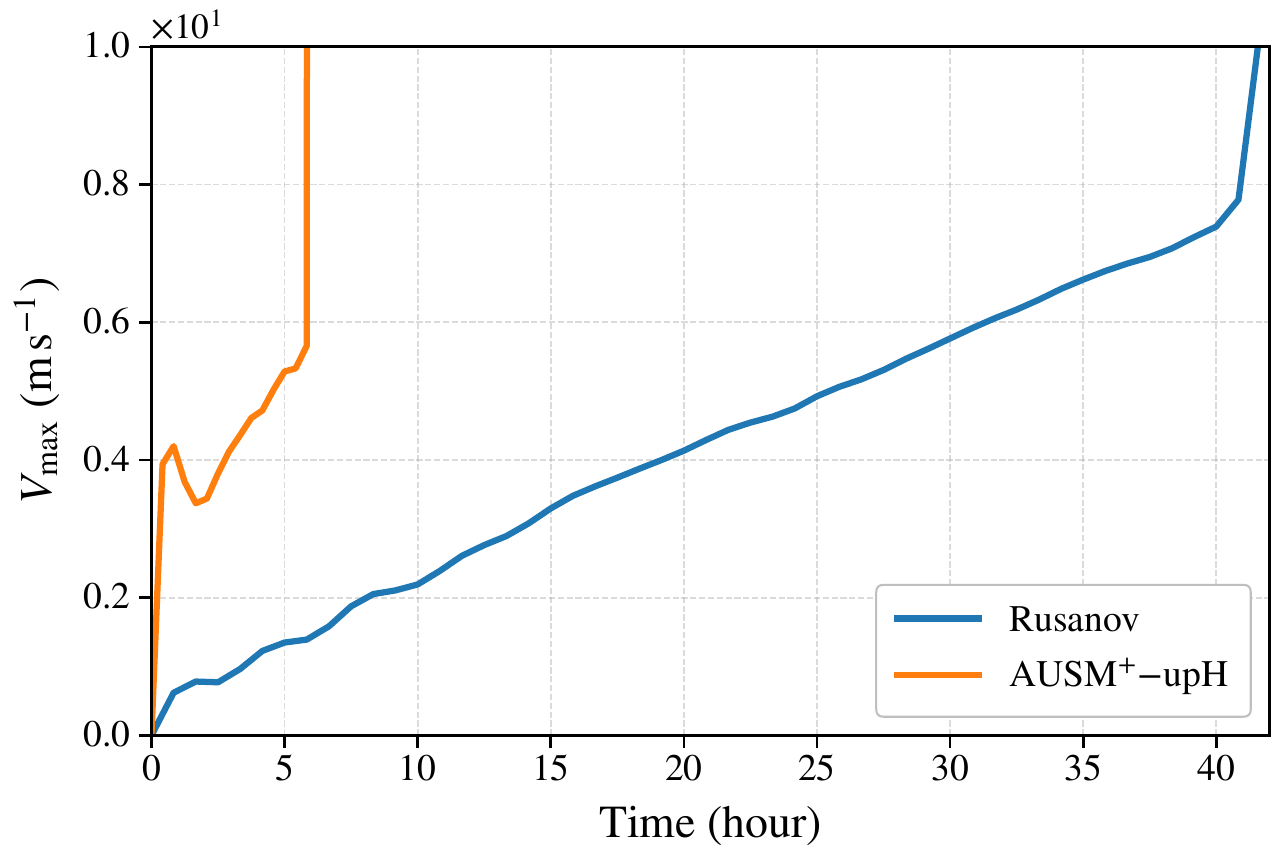}
        \caption{No deviation method}
        \label{fig:dcmip20_time_a}
    \end{subfigure}
    \hfill
    \begin{subfigure}[t]{0.48\textwidth}
        \centering
        \includegraphics[width=\textwidth]{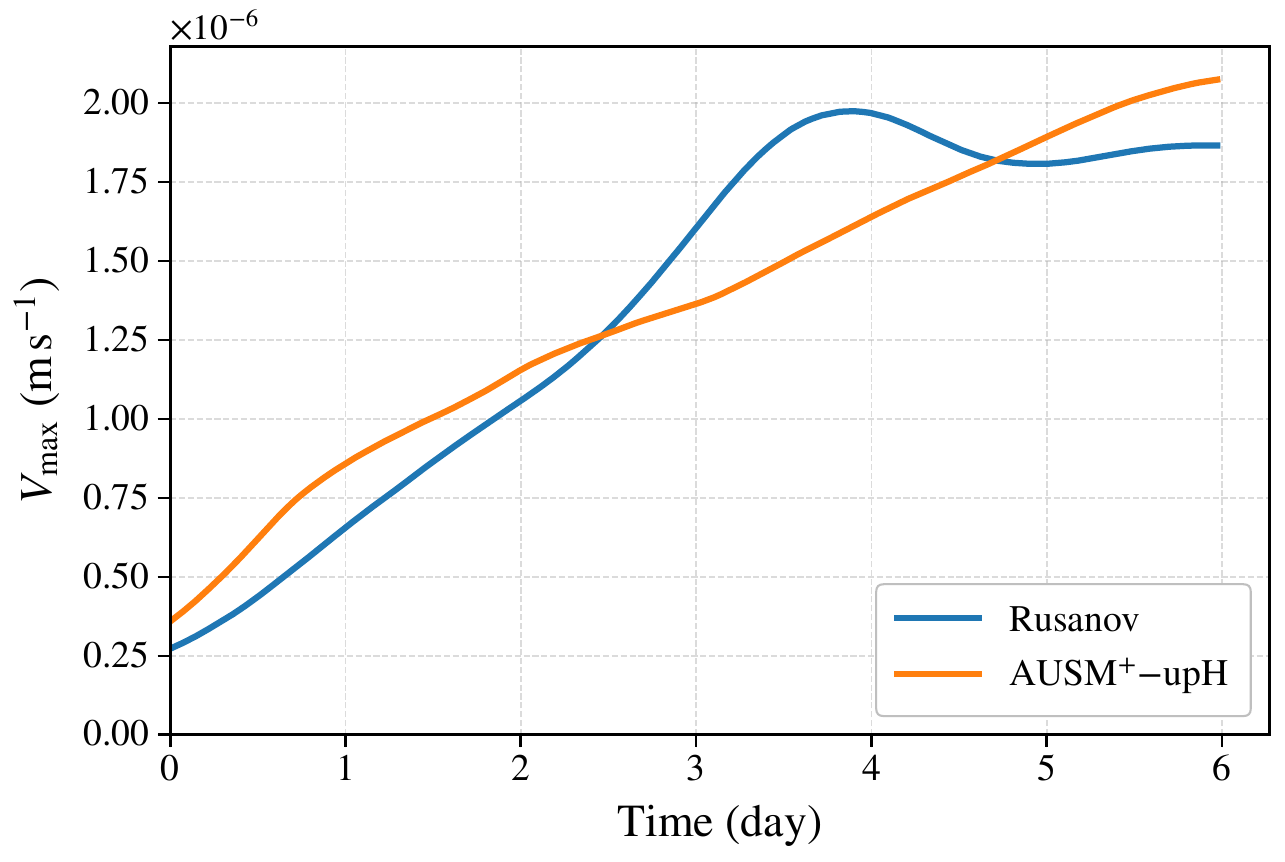}
        \caption{Deviation method}
        \label{fig:dcmip20_time_b}
    \end{subfigure}

    \caption{Time evolution of the maximum wind speed, $V_{\max}=\max_{\Omega}\sqrt{u^2+v^2+w^2}$, obtained using the Rusanov and AUSM$^{+}$-upH numerical fluxes. Panel (a) presents the simulations without the deviation formulation, and panel (b) the corresponding simulations with the deviation formulation. Note that the two panels use different scales on both axes. The curves in panel (b) have been smoothed for better visibility.}
    \label{fig:dcmip20_time}
\end{figure}

Figure~\ref{fig:dcmip20_time_a} shows that, in the absence of deviation formulation, both flux functions generate spurious velocities that grow continuously and ultimately lead to numerical failure. Therefore, neither simulation can be advanced to the six-day integration period recommended for this test case. For the Rusanov flux, $V_{\max}$ increases gradually during the initial stage of the simulation, indicating a progressive accumulation of numerical error, followed by a rapid increase after approximately 40~hours. In contrast, the AUSM$^{+}$-upH solution develops substantially larger errors at an earlier stage of the integration. This behaviour is consistent with the lower intrinsic dissipation of the AUSM$^{+}$-upH flux, which provides less damping of perturbations generated by the discrete hydrostatic imbalance.

The behaviour changes markedly when the deviation formulation is employed. As shown in Figure~\ref{fig:dcmip20_time_b}, the magnitude of the numerically generated velocity is substantially reduced, and both flux formulations remain stable throughout the full integration period. The initially nonzero wind speed is due to the added perturbation. Although the Rusanov and AUSM$^{+}$-upH solutions exhibit somewhat different temporal evolutions, the resulting values of $V_{\max}$ remain much smaller than in the non-well-balanced simulations and are of comparable magnitude ($\sim 10^{-6}\,\mathrm{m\,s^{-1}}$).

\begin{figure}[htbp]
    \centering

    \begin{subfigure}[t]{0.5\textwidth}
        \centering
        \includegraphics[width=\linewidth]{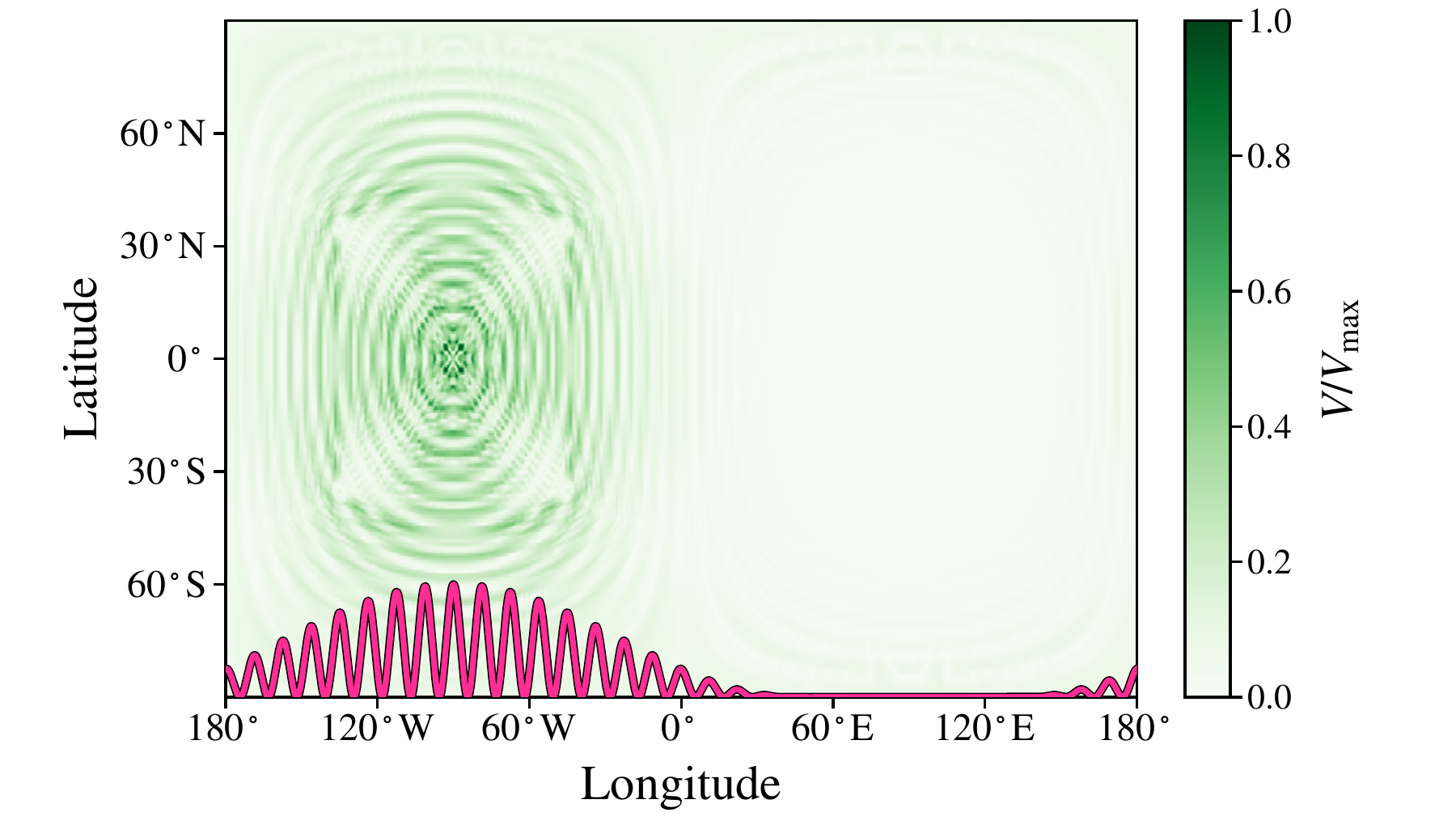}
        \caption{Rusanov (No DM)}
        \label{fig:dcmip20ca}
    \end{subfigure}%
    \begin{subfigure}[t]{0.5\textwidth}
        \centering
        \includegraphics[width=\linewidth]{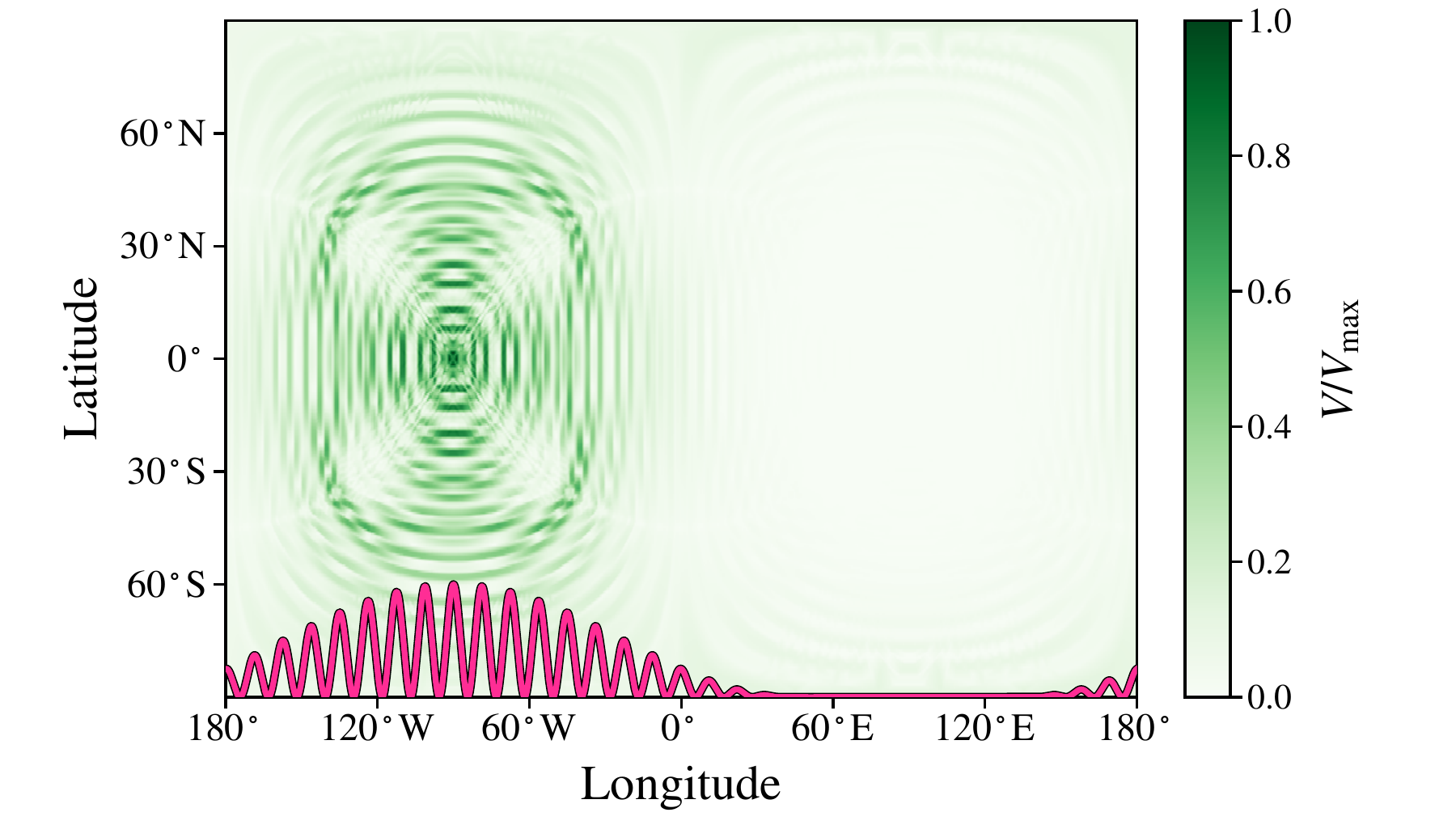}
        \caption{AUSM$^{+}$-upH (No DM)}
        \label{fig:dcmip20cb}
    \end{subfigure}

    \caption{Distribution of the normalized wind speed, $V/V_{\max}$, across the horizontal plane at the lowest model level at $t = 5~\mathrm{h}$. Panel (a) shows the Rusanov result without deviation method (DM), and panel (b) shows the corresponding AUSM$^{+}$-upH result.}
    \label{fig:dcmip20c}
\end{figure}

The spatial distribution of the numerical error in simulations without the deviation method is further illustrated in Figure~\ref{fig:dcmip20c}. For both numerical fluxes, the spurious flow initially develops in the vicinity of the mountainous region, with the largest velocities occurring near the mountain peak. As the simulation progresses, the error gradually propagates away from the topography and extends across the computational domain. It should be emphasized that the fields shown in Figure~\ref{fig:dcmip20c} are independently normalized by the maximum wind speed of each simulation. Consequently, the colour intensity characterizes the spatial structure and relative distribution of the spurious flow within each simulation, rather than its absolute magnitude.

\subsection{Non-Hydrostatic Gravity Waves}
\label{sec:gravitywave}
The nonhydrostatic gravity-wave test \citep{Ullrich2012DynamicalCM} evaluates the model's ability to accurately simulate short-timescale wave propagation, induced by a localized perturbation. The experiment is conducted on a non-rotating planet with a radius reduced by a factor of \(125\) relative to the Earth. The initial state is hydrostatically balanced and characterized by a constant Brunt-V\"ais\"al\"a frequency,
\[
N = 0.01~\mathrm{s}^{-1}.
\]
The background wind is
\[
u=u_0\cos\phi,
\qquad
v=0,
\qquad
w=0,
\]
with \(u_0=20~\mathrm{m\,s^{-1}}\).

The waves are generated by a localized potential-temperature perturbation,
\[
\Delta\theta(\lambda,\phi,z)
=
\Delta\theta_{\max}\,
\frac{d^2}{d^2+r^2}
\sin\left(\frac{2\pi z}{L_z}\right),
\]
where $\Delta\theta_{\max}=1~\mathrm{K}$, $L_z=20\,000~\mathrm{m}$ is the vertical wavelength, $d=5\,000~\mathrm{m}$ is the horizontal width parameter, and $r$ denotes the great-circle distance from the perturbation centre at
\[
(\lambda_c,\phi_c)
=
\left(\frac{2\pi}{3},0\right).
\]

The test uses \(N_e=16\) elements with \(N_s=5\) solution points per element in each horizontal direction, corresponding to a horizontal resolution of \(\Delta_h=1.125^\circ\). Vertically, two elements with five solution points per element provide 10 model levels up to a domain top of \(10\,000~\mathrm{m}\). Time integration is performed using a fixed time step of \(\Delta t=1~\mathrm{s}\). 

\begin{figure}[htbp]
    \centering

    \begin{subfigure}[t]{0.48\textwidth}
        \centering
        \includegraphics[width=\textwidth]{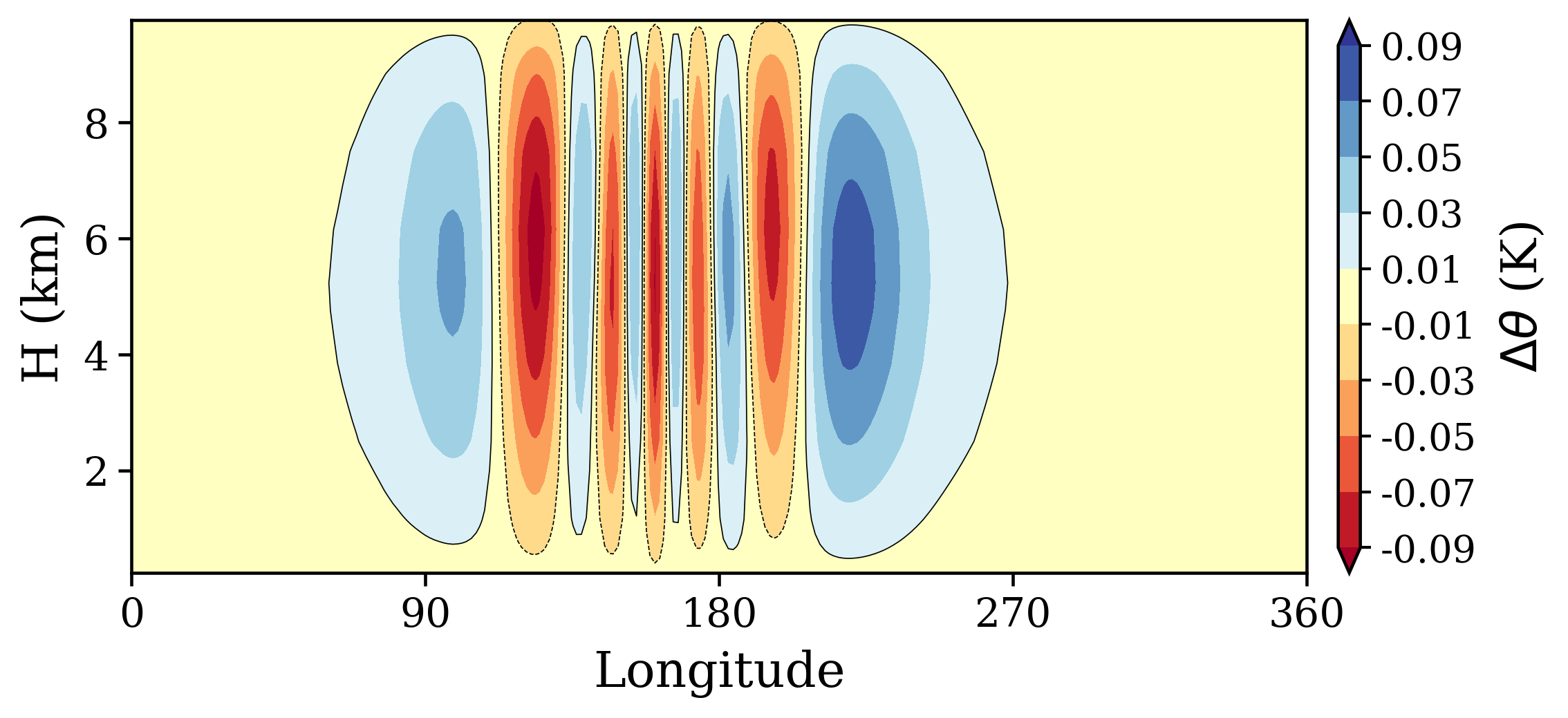}
        \caption{Rusanov (No DM)}
        \label{fig:dcmip31a}
    \end{subfigure}
    \hfill
    \begin{subfigure}[t]{0.48\textwidth}
        \centering
        \includegraphics[width=\textwidth]{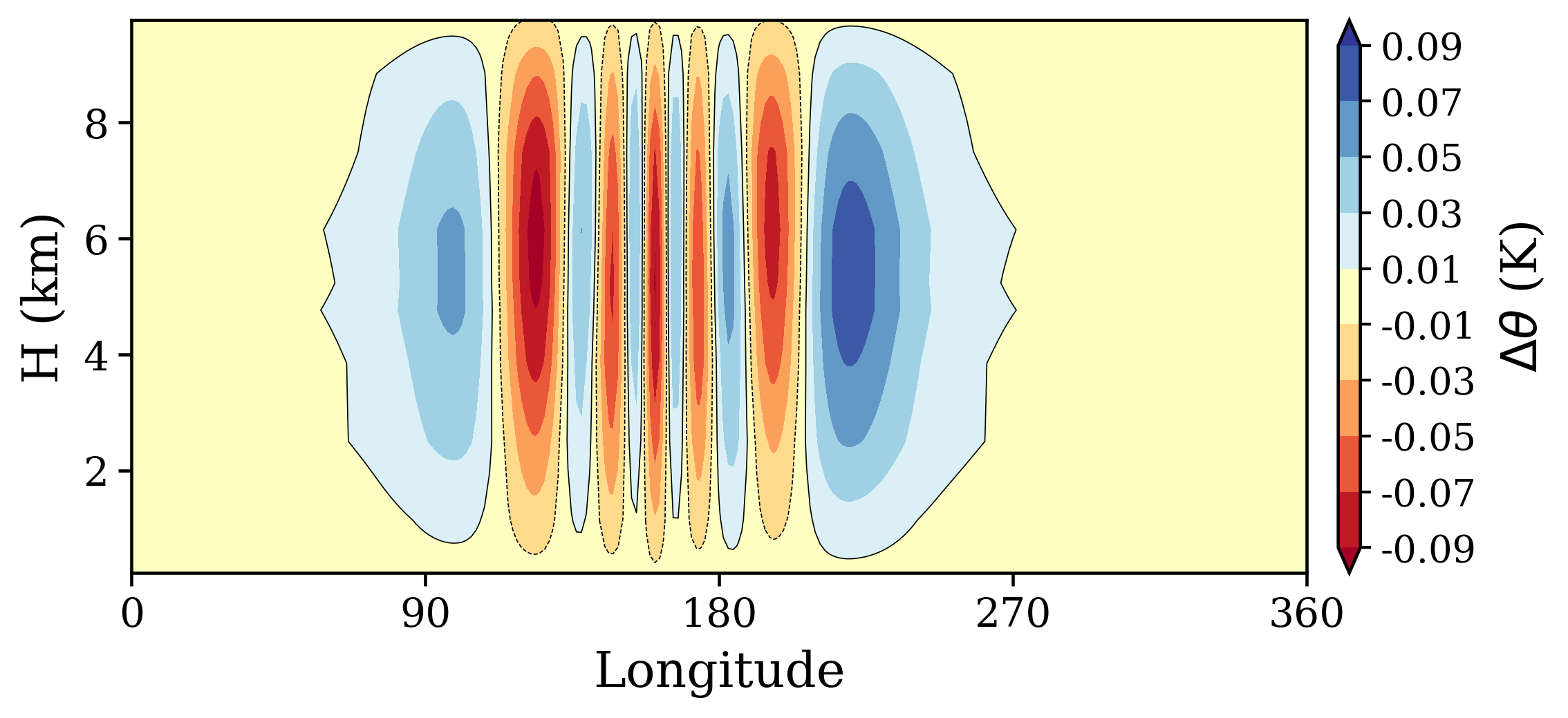}
        \caption{AUSM$^+$-upH (No DM)}
        \label{fig:dcmip31b}
    \end{subfigure}

    \vspace{0.3cm}

    \begin{subfigure}[t]{0.48\textwidth}
        \centering
        \includegraphics[width=\textwidth]{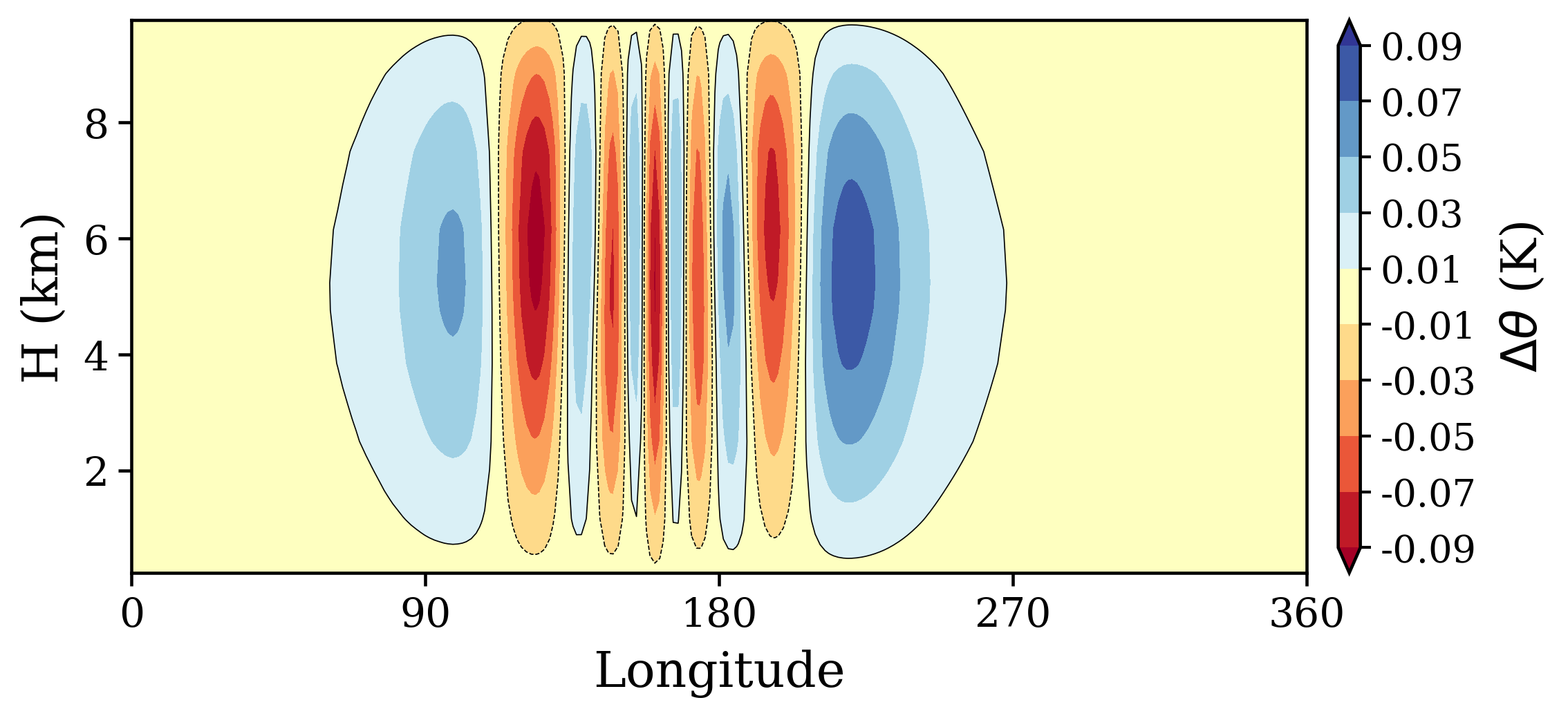}
        \caption{Rusanov (DM)}
        \label{fig:dcmip31c}
    \end{subfigure}
    \hfill
    \begin{subfigure}[t]{0.48\textwidth}
        \centering
        \includegraphics[width=\textwidth]{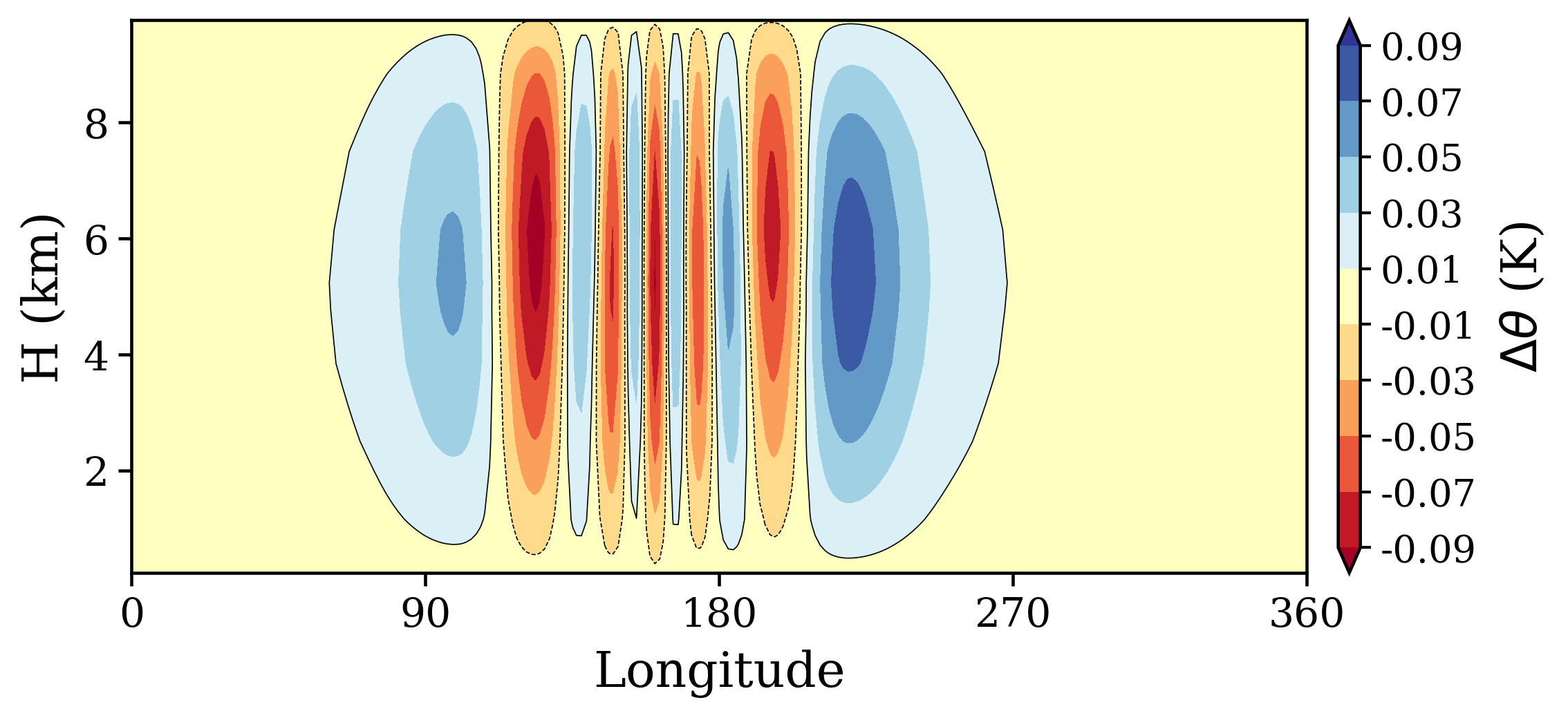}
        \caption{AUSM$^+$-upH (DM)}
        \label{fig:dcmip31d}
    \end{subfigure}

    \caption{Potential-temperature perturbation $\Delta\theta$ along the equator for the non-hydrostatic gravity-wave test at \(t=1800~\mathrm{s}\). DM stands for the deviation method.}
    \label{fig:dcmip31}
\end{figure}

In Figure~\ref{fig:dcmip31}, two-dimensional equatorial cross-sections of $\Delta\theta$ at $t=1800~\mathrm{s}$ illustrate the structure of the propagating wave packet throughout the $10~\mathrm{km}$ vertical domain. Without the deviation method, the low-dissipation AUSM$^+$-upH flux (Figure~\ref{fig:dcmip31b}) exhibits noticeable jagged contours near the outer edges of the wave train (around longitudes $90^\circ$ and $250^\circ$). These artefacts stem from discretization errors introduced when evaluating the small perturbations in the presence of a much larger hydrostatic background state. In contrast, the Rusanov flux without the deviation formulation (Figure~\ref{fig:dcmip31a}) produces smoother contours, as its greater intrinsic numerical dissipation suppresses hydrostatic imbalance errors along with high-frequency modes. When the deviation well-balanced formulation is enabled (Figures~\ref{fig:dcmip31c} and \ref{fig:dcmip31d}), the background hydrostatic state is preserved more accurately, eliminating the spurious background noise in AUSM$^+$-upH and producing smooth wave contours for both flux functions.

\begin{figure}[ht]
    \centering

    \begin{subfigure}[t]{0.95\textwidth}
        \centering
        \includegraphics[width=\textwidth]{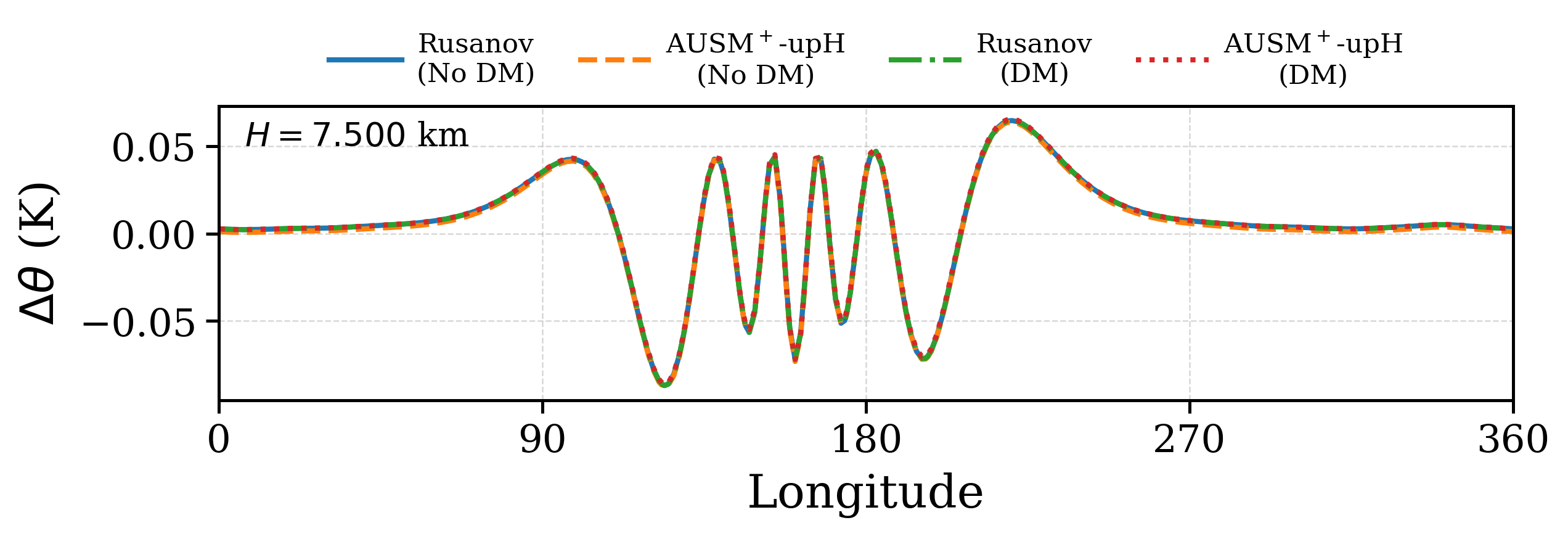}
        \caption{Full longitudinal profile.}
        \label{fig:dcmip31_line_full}
    \end{subfigure}

    \vspace{0.3cm}

    \begin{subfigure}[t]{0.95\textwidth}
        \centering
        \includegraphics[width=\textwidth]{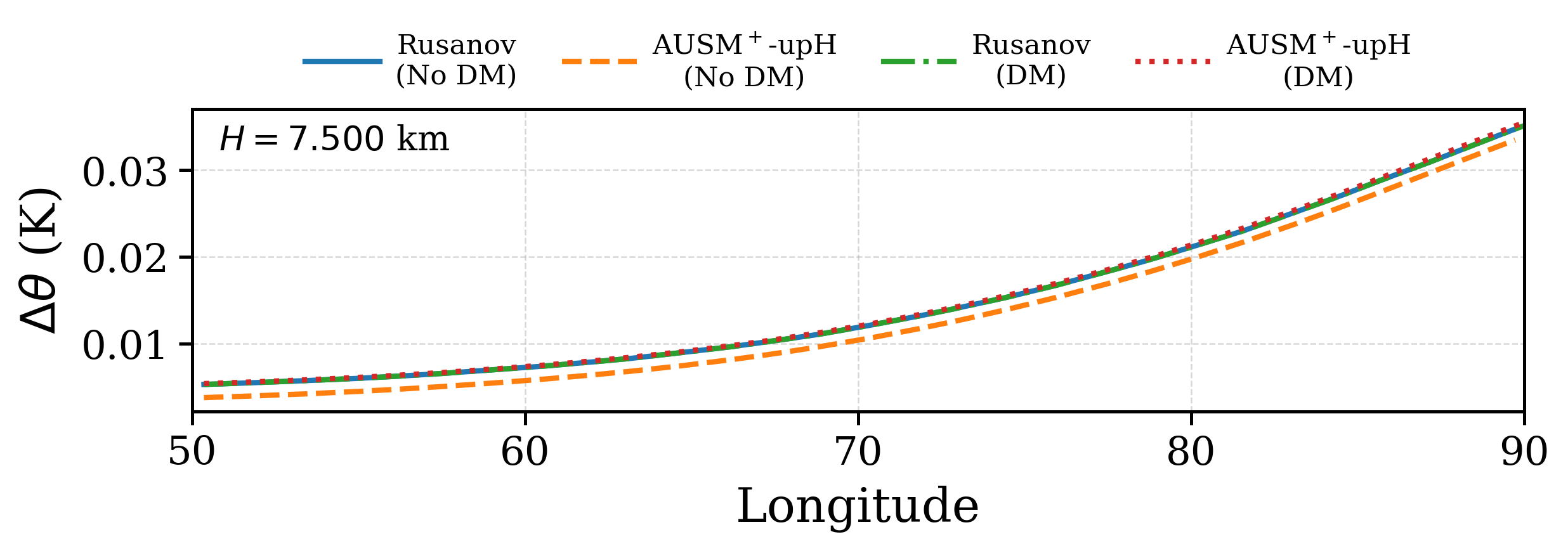}
        \caption{Zoomed view of the left outer-lobe region.}
        \label{fig:dcmip31_line_zoom}
    \end{subfigure}

    \caption{Potential-temperature perturbation $\Delta\theta$ along the equator at $7500~\mathrm{m}$ and $t=1800~\mathrm{s}$ for the Rusanov and AUSM$^+$-upH fluxes, with and without the deviation formulation.}
    \label{fig:dcmip31_line}
\end{figure}

Figure~\ref{fig:dcmip31_line} shows the one-dimensional potential-temperature perturbation profiles at \(H=7.500~\mathrm{km}\) and \(t=1800~\mathrm{s}\) for all four configurations. Over the full longitude range, the solutions look nearly indistinguishable, indicating that the overall phase and amplitude of the gravity-wave packet are similarly reproduced by all configurations. However, the zoomed view of the left outer-lobe region shows a slight departure for the non-well-balanced AUSM$^+$-upH solution from others, confirming the results of Figure~\ref{fig:dcmip31b}. A similar behaviour is also observed in the right outer lobe.

\section{Conclusion} \label{summary}

This work investigated a high-order DFR framework coupled with exponential time integration for simulating low-Mach-number, nearly hydrostatic atmospheric flows governed by the 3D compressible Euler equations with gravity. The AUSM$^+$-upH flux was employed to provide a more appropriate low-Mach scaling of the numerical dissipation, while the deviation well-balanced formulation was used to improve the discrete preservation of hydrostatically balanced states.

The Gresho-vortex experiments demonstrate the importance of an appropriate low-Mach numerical flux. As the Mach number decreases, the Rusanov flux becomes increasingly diffusive and eventually develops spurious structures, whereas AUSM$^+$-upH preserves the vortex much more effectively. Despite this improved low-Mach behaviour, the two-dimensional isothermal hydrostatic-equilibrium test shows that AUSM$^+$-upH alone is insufficient for exactly or nearly hydrostatically balanced states. In such cases, the deviation method is necessary to accurately preserve the hydrostatic balance. Small residual imbalances in the discrete balance law can be amplified by AUSM$^+$-upH, whereas the inherent numerical diffusion of the Rusanov flux tends to damp these perturbations.

The rising thermal-bubble experiment represents a different regime, in which the physical perturbation is sufficiently strong that the deviation formulation has only a limited influence on the resolved dynamics. In this case, the dominant differences arise from the choice of numerical flux. AUSM$^+$-upH preserves a sharper thermal structure, whereas the greater dissipation of the Rusanov flux produces a more smeared response.

The global steady-state atmosphere over orography provides a more stringent assessment of hydrostatic preservation. Without the deviation method, both numerical fluxes generate spurious velocities that grow in time, and neither configuration can complete the intended integration. With the deviation formulation, these errors are reduced by several orders of magnitude, allowing both fluxes to remain stable throughout the full integration period.

In the non-hydrostatic gravity-wave experiment, the different configurations produce similar results in terms of the propagation, phase, and amplitude of the wave packet. However, the AUSM$^+$-upH solution without the deviation formulation develops small-scale artefacts. These features are attributed to discretization errors arising when small dynamical perturbations are evaluated against a much larger hydrostatic background state. The more dissipative Rusanov flux suppresses these artefacts, whereas the deviation formulation reduces the underlying hydrostatic imbalance, allowing the low-dissipation AUSM$^+$-upH flux to capture the wave structure without relying on excessive numerical diffusion.

These results highlight an important interaction between low dissipation and well balancing. Reducing numerical diffusion is desirable for preserving weak atmospheric motions, but it also exposes discretization errors that may otherwise be masked by a more dissipative flux. A well-balanced treatment is therefore particularly important when low-dissipation numerical fluxes are applied to nearly hydrostatic atmospheric flows. Low-Mach correction and well balancing should consequently be regarded as complementary components of a compressible atmospheric discretization. Combined with high-order DFR spatial discretization and exponential time integration, these techniques provide a promising framework for accurately simulating compressible atmospheric flows containing both fast acoustic modes and slowly evolving, nearly balanced dynamics.

\section*{Acknowledgements}

The authors thank Philipp Edelmann for valuable discussions, insightful comments, and helpful suggestions that contributed to the development of this work.

\section*{Code Availability}

The source code used for the numerical experiments presented in this study is publicly available at \url{https://github.com/Wx-Alliance-Alliance-Meteo/WxFactory}.

\bibliography{references}

\begin{thebibliography}{51}
\providecommand{\natexlab}[1]{#1}
\providecommand{\url}[1]{\texttt{#1}}
\expandafter\ifx\csname urlstyle\endcsname\relax
  \providecommand{\doi}[1]{doi: #1}\else
  \providecommand{\doi}{doi: \begingroup \urlstyle{rm}\Url}\fi

\bibitem[Artiano et~al.(2026)Artiano, Knoth, Spichtinger, and Ranocha]{ARTIANO2026115221}
Marco Artiano, Oswald Knoth, Peter Spichtinger, and Hendrik Ranocha.
\newblock Structure-preserving high-order methods for the compressible euler equations in potential temperature formulation for atmospheric flows.
\newblock \emph{Journal of Computational Physics}, 566:\penalty0 115221, 2026.
\newblock ISSN 0021-9991.
\newblock \doi{https://doi.org/10.1016/j.jcp.2026.115221}.
\newblock URL \url{https://www.sciencedirect.com/science/article/pii/S0021999126005723}.

\bibitem[Barsukow(2025)]{barsukow-2025}
Wasilij Barsukow.
\newblock {Stationarity preservation and the low Mach number behaviour of the Discontinuous Galerkin method on Cartesian grids}.
\newblock \emph{arXiv (Cornell University)}, 11 2025.
\newblock \doi{10.48550/arxiv.2511.18505}.
\newblock URL \url{https://doi.org/10.48550/arxiv.2511.18505}.

\bibitem[Berberich et~al.(2021)Berberich, Chandrashekar, and Klingenberg]{Berberich_Chandrashekar_Klingenberg_2021}
Jonas~P. Berberich, Praveen Chandrashekar, and Christian Klingenberg.
\newblock High order well-balanced finite volume methods for multi-dimensional systems of hyperbolic balance laws.
\newblock \emph{Computers \& Fluids}, 219:\penalty0 104858, Apr 2021.
\newblock \doi{10.1016/j.compfluid.2021.104858}.

\bibitem[Birken(2021)]{birken:21}
Philipp Birken.
\newblock \emph{{Numerical Methods for Unsteady Compressible Flow Problems}}.
\newblock CRC Press, 2021.

\bibitem[Birken et~al.(2026)Birken, Dedner, and Klöfkorn]{bidekl:26}
Philipp Birken, Andreas Dedner, and Robert Klöfkorn.
\newblock A jacobian-free multigrid preconditioner for discontinuous galerkin methods applied to atmospheric flows.
\newblock \emph{GEM - Int. J. Geomath.}, 17, 12 2026.
\newblock ISSN 18692680.
\newblock \doi{10.1007/s13137-025-00279-y}.

\bibitem[Botta et~al.(2004)Botta, Klein, Langenberg, and Lützenkirchen]{BOTTA2004539}
Nicola Botta, Rupert Klein, Susanne Langenberg, and Susanne Lützenkirchen.
\newblock Well balanced finite volume methods for nearly hydrostatic flows.
\newblock \emph{Journal of Computational Physics}, 196\penalty0 (2):\penalty0 539--565, 2004.
\newblock ISSN 0021-9991.
\newblock \doi{https://doi.org/10.1016/j.jcp.2003.11.008}.
\newblock URL \url{https://www.sciencedirect.com/science/article/pii/S0021999103006077}.

\bibitem[Cargo and Roux(1994)]{Cargo1994UnS}
P.~Cargo and Alain~Le Roux.
\newblock Un sch{\'e}ma {\'e}quilibre adapt{\'e} au mod{\`e}le d'atmosph{\`e}re avec termes de gravit{\'e}.
\newblock 1994.
\newblock URL \url{https://api.semanticscholar.org/CorpusID:171183211}.

\bibitem[Chertock et~al.(2023)Chertock, Kurganov, Wu, and Yan]{chertock2023}
Alina Chertock, Alexander Kurganov, Tong Wu, and Jun Yan.
\newblock Well-balanced numerical method for atmospheric flow equations with gravity.
\newblock \emph{Appl. Math. Comput.}, 439\penalty0 (C), February 2023.
\newblock ISSN 0096-3003.
\newblock \doi{10.1016/j.amc.2022.127587}.
\newblock URL \url{https://doi.org/10.1016/j.amc.2022.127587}.

\bibitem[Clancy and Pudykiewicz(2013)]{Clancy01122013}
Colm Clancy and Janusz~A. Pudykiewicz.
\newblock On the use of exponential time integration methods in atmospheric models.
\newblock \emph{Tellus A: Dynamic Meteorology and Oceanography}, 65\penalty0 (1):\penalty0 20898, 2013.
\newblock \doi{10.3402/tellusa.v65i0.20898}.

\bibitem[Clark(1977)]{Clark1977}
Terry~L. Clark.
\newblock A small-scale dynamic model using a terrain-following coordinate transformation.
\newblock \emph{Journal of Computational Physics}, 24\penalty0 (2):\penalty0 186--215, 1977.
\newblock \doi{10.1016/0021-9991(77)90057-2}.

\bibitem[Cordier et~al.(2012)Cordier, Degond, and Kumbaro]{CORDIER20125685}
Floraine Cordier, Pierre Degond, and Anela Kumbaro.
\newblock An asymptotic-preserving all-speed scheme for the euler and navier–stokes equations.
\newblock \emph{Journal of Computational Physics}, 231\penalty0 (17):\penalty0 5685--5704, 2012.
\newblock ISSN 0021-9991.
\newblock \doi{https://doi.org/10.1016/j.jcp.2012.04.025}.
\newblock URL \url{https://www.sciencedirect.com/science/article/pii/S0021999112002069}.

\bibitem[Dedner(2003)]{dedner:04}
Andreas Dedner.
\newblock \emph{Solving the system of Radiation Magnetohydrodynamics for solar physical simulations in 3d}.
\newblock PhD thesis, Universität Freiburg, 2003.

\bibitem[{Edelmann, Philipp V. F.} et~al.(2021){Edelmann, Philipp V. F.}, {Horst, L.}, {Berberich, Jonas P.}, {Andrassy, R.}, {Higl, Johann}, {Leidi, Giovanni}, {Klingenberg, Christian}, and {Röpke, Friedrich K.}]{Edelmann2021}
{Edelmann, Philipp V. F.}, {Horst, L.}, {Berberich, Jonas P.}, {Andrassy, R.}, {Higl, Johann}, {Leidi, Giovanni}, {Klingenberg, Christian}, and {Röpke, Friedrich K.}
\newblock Well-balanced treatment of gravity in astrophysical fluid dynamics simulations at low mach numbers.
\newblock \emph{A\&A}, 652:\penalty0 A53, 2021.
\newblock \doi{10.1051/0004-6361/202140653}.
\newblock URL \url{https://doi.org/10.1051/0004-6361/202140653}.

\bibitem[Gassner and Kopriva(2011)]{gassner2011comparison}
Gregor Gassner and David~A Kopriva.
\newblock A comparison of the dispersion and dissipation errors of {Gauss} and {Gauss}--{Lobatto} discontinuous {Galerkin} spectral element methods.
\newblock \emph{SIAM Journal on Scientific Computing}, 33\penalty0 (5):\penalty0 2560--2579, 2011.

\bibitem[Gaudreault et~al.(2022)Gaudreault, Charron, Dallerit, and Tokman]{gaudreault2022high}
St{\'e}phane Gaudreault, Martin Charron, Valentin Dallerit, and Mayya Tokman.
\newblock High-order numerical solutions to the shallow-water equations on the rotated cubed-sphere grid.
\newblock \emph{Journal of Computational Physics}, 449:\penalty0 110792, 2022.

\bibitem[Gaudreault et~al.(2026)Gaudreault, Subich, Panday, Charron, Magnoux, Dallerit, and Tokman]{gaudreault2026application}
St{\'e}phane Gaudreault, Christopher Subich, Shoyon Panday, Martin Charron, Vincent Magnoux, Valentin Dallerit, and Mayya Tokman.
\newblock Application of high-order direct flux reconstruction and stiffness-resilient time integration to simulations of idealized atmospheric flows.
\newblock \emph{International Journal for Numerical Methods in Fluids}, 98\penalty0 (4):\penalty0 448--468, 2026.

\bibitem[Gaudreault et~al.(2018)Gaudreault, Rainwater, and Tokman]{GAUDREAULT2018236}
Stéphane Gaudreault, Greg Rainwater, and Mayya Tokman.
\newblock Kiops: A fast adaptive krylov subspace solver for exponential integrators.
\newblock \emph{Journal of Computational Physics}, 372:\penalty0 236--255, 2018.
\newblock ISSN 0021-9991.
\newblock \doi{https://doi.org/10.1016/j.jcp.2018.06.026}.
\newblock URL \url{https://www.sciencedirect.com/science/article/pii/S0021999118304042}.

\bibitem[Giraldo and Restelli(2008)]{GIRALDO20083849}
Francis~X. Giraldo and Marco Restelli.
\newblock A study of spectral element and discontinuous galerkin methods for the navier–stokes equations in nonhydrostatic mesoscale atmospheric modeling: Equation sets and test cases.
\newblock \emph{Journal of Computational Physics}, 227\penalty0 (8):\penalty0 3849--3877, 2008.
\newblock ISSN 0021-9991.
\newblock \doi{https://doi.org/10.1016/j.jcp.2007.12.009}.
\newblock URL \url{https://www.sciencedirect.com/science/article/pii/S0021999107005384}.

\bibitem[Gouasmi et~al.(2022)Gouasmi, Murman, and Duraisamy]{GOUASMI2022111036}
Ayoub Gouasmi, Scott~M. Murman, and Karthik Duraisamy.
\newblock Entropy-stable schemes in the low-mach-number regime: Flux-preconditioning, entropy breakdowns, and entropy transfers.
\newblock \emph{Journal of Computational Physics}, 456:\penalty0 111036, 2022.
\newblock ISSN 0021-9991.
\newblock \doi{https://doi.org/10.1016/j.jcp.2022.111036}.
\newblock URL \url{https://www.sciencedirect.com/science/article/pii/S0021999122000985}.

\bibitem[Guillard and Nkonga(2017)]{GuillardNkonga2017}
Herv{\'e} Guillard and Boniface Nkonga.
\newblock On the behaviour of upwind schemes in the low mach number limit: A review.
\newblock In \emph{Handbook of Numerical Methods for Hyperbolic Problems: Applied and Modern Issues}, volume~18 of \emph{Handbook of Numerical Analysis}, pages 203--231. Elsevier/North-Holland, Amsterdam, 2017.
\newblock \doi{10.1016/bs.hna.2016.09.002}.

\bibitem[Guillard and Viozat(1999)]{Guillard1999OnTB}
Herv{\'e} Guillard and C{\'e}cile Viozat.
\newblock On the behaviour of upwind schemes in the low mach number limit.
\newblock \emph{Computers \& Fluids}, 28:\penalty0 63--86, 1999.
\newblock URL \url{https://api.semanticscholar.org/CorpusID:122028682}.

\bibitem[Huynh(2007)]{huynh2007flux}
Hung~T Huynh.
\newblock A flux reconstruction approach to high-order schemes including discontinuous {Galerkin} methods.
\newblock In \emph{18th AIAA Computational Fluid Dynamics Conference}, page 4079, 2007.

\bibitem[Jung and Perrier(2024)]{Jung2024}
Jonathan Jung and Vincent Perrier.
\newblock Behavior of the discontinuous galerkin method for compressible flows at low mach number on triangles and tetrahedrons.
\newblock \emph{SIAM Journal on Scientific Computing}, 46\penalty0 (1):\penalty0 A452--A482, 2024.
\newblock \doi{10.1137/23M154755X}.
\newblock URL \url{https://doi.org/10.1137/23M154755X}.

\bibitem[K{\"a}ppeli(2022)]{Kappeli2022}
Roger K{\"a}ppeli.
\newblock Well-balanced methods for computational astrophysics.
\newblock \emph{Living Reviews in Computational Astrophysics}, 8\penalty0 (2), 2022.
\newblock \doi{10.1007/s41115-022-00014-6}.

\bibitem[Kelly and Giraldo(2012)]{KELLY20127988}
James~F. Kelly and Francis~X. Giraldo.
\newblock Continuous and discontinuous galerkin methods for a scalable three-dimensional nonhydrostatic atmospheric model: Limited-area mode.
\newblock \emph{Journal of Computational Physics}, 231\penalty0 (24):\penalty0 7988--8008, 2012.
\newblock ISSN 0021-9991.
\newblock \doi{https://doi.org/10.1016/j.jcp.2012.04.042}.
\newblock URL \url{https://www.sciencedirect.com/science/article/pii/S0021999112002252}.

\bibitem[Liou(2006)]{LIOU2006137}
Meng-Sing Liou.
\newblock A sequel to ausm, part ii: Ausm+-up for all speeds.
\newblock \emph{Journal of Computational Physics}, 214\penalty0 (1):\penalty0 137--170, 2006.
\newblock ISSN 0021-9991.
\newblock \doi{https://doi.org/10.1016/j.jcp.2005.09.020}.
\newblock URL \url{https://www.sciencedirect.com/science/article/pii/S0021999105004274}.

\bibitem[Liska and Wendroff(2003)]{liska2003comparison}
Richard Liska and Burton Wendroff.
\newblock Comparison of several difference schemes on 1d and 2d test problems for the euler equations.
\newblock \emph{SIAM Journal on Scientific Computing}, 25\penalty0 (3):\penalty0 995--1017, 2003.
\newblock \doi{10.1137/S1064827502402120}.

\bibitem[Liu et~al.(2025)Liu, Guo, Jiang, and Zhang]{ligujizh:25}
Yuchang Liu, Wei Guo, Yan Jiang, and Mengping Zhang.
\newblock {Structure-preserving nodal DG method for the Euler equations with gravity: well-balanced, entropy stable, and positivity preserving}.
\newblock \emph{J. Comp. Phys.}, 537, 9 2025.
\newblock ISSN 10902716.
\newblock \doi{10.1016/j.jcp.2025.114095}.

\bibitem[Marras et~al.(2016)Marras, Kelly, Moragues, M{\"u}ller, Kopera, V{\'a}zquez, Giraldo, Houzeaux, and Jorba]{Marras2016}
Simone Marras, James~F. Kelly, Maria Moragues, Andreas M{\"u}ller, Michal~A. Kopera, Mariano V{\'a}zquez, Francis~X. Giraldo, Guillaume Houzeaux, and Oriol Jorba.
\newblock A review of element-based galerkin methods for numerical weather prediction: Finite elements, spectral elements, and discontinuous galerkin.
\newblock \emph{Archives of Computational Methods in Engineering}, 23:\penalty0 673--722, 2016.
\newblock \doi{10.1007/s11831-015-9152-1}.

\bibitem[Mengaldo et~al.(2015)Mengaldo, De~Grazia, Moxey, Vincent, and Sherwin]{Mengaldo2015}
Gianmarco Mengaldo, Daniele De~Grazia, David Moxey, Peter Vincent, and Spencer~J. Sherwin.
\newblock Dealiasing techniques for high-order spectral element methods on regular and irregular grids.
\newblock \emph{Journal of Computational Physics}, 299:\penalty0 56--81, 2015.
\newblock \doi{10.1016/j.jcp.2015.06.032}.

\bibitem[Miczek(2013)]{Miczek2013}
Florian Miczek.
\newblock \emph{Simulation of Low Mach Number Astrophysical Flows}.
\newblock Ph.d. thesis, Technische Universit{\"a}t M{\"u}nchen, Munich, Germany, 2013.

\bibitem[Miczek et~al.(2015)Miczek, R{\"o}pke, and Edelmann]{Miczek2015}
Florian Miczek, Friedrich~K. R{\"o}pke, and Philipp V.~F. Edelmann.
\newblock New numerical solver for flows at various mach numbers.
\newblock \emph{Astronomy \& Astrophysics}, 576:\penalty0 A50, 2015.
\newblock \doi{10.1051/0004-6361/201425059}.

\bibitem[Oßwald et~al.(2016)Oßwald, Siegmund, Birken, Hannemann, and Meister]{L2ROE}
K.~Oßwald, A.~Siegmund, Philipp Birken, V.~Hannemann, and Andreas Meister.
\newblock L2roe: a low dissipation version of roe's approximate riemann solver for low mach numbers.
\newblock \emph{International Journal for Numerical Methods in Fluids}, 81\penalty0 (2):\penalty0 71--86, 2016.
\newblock \doi{https://doi.org/10.1002/fld.4175}.
\newblock URL \url{https://onlinelibrary.wiley.com/doi/abs/10.1002/fld.4175}.

\bibitem[Persson and Peraire(2006)]{Persson}
Per-Olof Persson and Jaime Peraire.
\newblock \emph{Sub-Cell Shock Capturing for Discontinuous Galerkin Methods}.
\newblock 2006.
\newblock \doi{10.2514/6.2006-112}.
\newblock URL \url{https://arc.aiaa.org/doi/abs/10.2514/6.2006-112}.

\bibitem[Rieper(2011)]{RIEPER20115263}
Felix Rieper.
\newblock A low-mach number fix for roe’s approximate riemann solver.
\newblock \emph{Journal of Computational Physics}, 230\penalty0 (13):\penalty0 5263--5287, 2011.
\newblock ISSN 0021-9991.
\newblock \doi{https://doi.org/10.1016/j.jcp.2011.03.025}.
\newblock URL \url{https://www.sciencedirect.com/science/article/pii/S0021999111001689}.

\bibitem[Robert(1993)]{robert1993bubble}
Andr{\'e} Robert.
\newblock Bubble convection experiments with a semi-implicit formulation of the {Euler} equations.
\newblock \emph{Journal of the Atmospheric Sciences}, 50\penalty0 (13):\penalty0 1865--1873, 1993.

\bibitem[Romero(2017)]{romero2017development}
Joshua Romero.
\newblock \emph{On the development of the direct flux reconstruction scheme for high-order fluid flow simulations}.
\newblock PhD thesis, Stanford University, 2017.

\bibitem[Romero et~al.(2016)Romero, Asthana, and Jameson]{romero2016simplified}
Joshua Romero, Kartikey Asthana, and Antony Jameson.
\newblock A simplified formulation of the flux reconstruction method.
\newblock \emph{Journal of Scientific Computing}, 67\penalty0 (1):\penalty0 351--374, 2016.

\bibitem[Rusanov(1962)]{RUSANOV1962304}
V.V Rusanov.
\newblock The calculation of the interaction of non-stationary shock waves and obstacles.
\newblock \emph{USSR Computational Mathematics and Mathematical Physics}, 1\penalty0 (2):\penalty0 304--320, 1962.
\newblock ISSN 0041-5553.
\newblock \doi{https://doi.org/10.1016/0041-5553(62)90062-9}.
\newblock URL \url{https://www.sciencedirect.com/science/article/pii/0041555362900629}.

\bibitem[Sadourny(1972)]{Sadourny1972}
Robert Sadourny.
\newblock Conservative finite-difference approximations of the primitive equations on quasi-uniform spherical grids.
\newblock \emph{Monthly Weather Review}, 100:\penalty0 136--144, 1972.

\bibitem[Schwarz et~al.(2025)Schwarz, Kempf, Keim, Kopper, Rohde, and Beck]{SCHWARZ2025106874}
Anna Schwarz, Daniel Kempf, Jens Keim, Patrick Kopper, Christian Rohde, and Andrea Beck.
\newblock Comparison of entropy stable collocation high-order dg methods for compressible turbulent flows.
\newblock \emph{Computers \& Fluids}, 303:\penalty0 106874, 2025.
\newblock ISSN 0045-7930.
\newblock \doi{https://doi.org/10.1016/j.compfluid.2025.106874}.
\newblock URL \url{https://www.sciencedirect.com/science/article/pii/S0045793025003342}.

\bibitem[song Li and wei Gu(2008)]{LI20085144}
Xue song Li and Chun wei Gu.
\newblock An all-speed roe-type scheme and its asymptotic analysis of low mach number behaviour.
\newblock \emph{Journal of Computational Physics}, 227\penalty0 (10):\penalty0 5144--5159, 2008.
\newblock ISSN 0021-9991.
\newblock \doi{https://doi.org/10.1016/j.jcp.2008.01.037}.
\newblock URL \url{https://www.sciencedirect.com/science/article/pii/S0021999108000697}.

\bibitem[Thornber et~al.(2008)Thornber, Mosedale, Drikakis, Youngs, and Williams]{THORNBER20084873}
Ben Thornber, A.~Mosedale, Dimitris Drikakis, David Youngs, and R.J.R. Williams.
\newblock An improved reconstruction method for compressible flows with low mach number features.
\newblock \emph{Journal of Computational Physics}, 227\penalty0 (10):\penalty0 4873--4894, 2008.
\newblock ISSN 0021-9991.
\newblock \doi{https://doi.org/10.1016/j.jcp.2008.01.036}.
\newblock URL \url{https://www.sciencedirect.com/science/article/pii/S0021999108000429}.

\bibitem[Tokman(2006)]{TOKMAN2006748}
Mayya Tokman.
\newblock Efficient integration of large stiff systems of odes with exponential propagation iterative (epi) methods.
\newblock \emph{Journal of Computational Physics}, 213\penalty0 (2):\penalty0 748--776, 2006.
\newblock ISSN 0021-9991.
\newblock \doi{https://doi.org/10.1016/j.jcp.2005.08.032}.
\newblock URL \url{https://www.sciencedirect.com/science/article/pii/S0021999105004158}.

\bibitem[Turkel(1999)]{Turkel1999}
Eli Turkel.
\newblock Preconditioning techniques in computational fluid dynamics.
\newblock \emph{Annual Review of Fluid Mechanics}, 31\penalty0 (1):\penalty0 385--416, 1999.
\newblock \doi{10.1146/annurev.fluid.31.1.385}.

\bibitem[Ullrich et~al.(2012)Ullrich, Jablonowski, Kent, Lauritzen, and Nair]{Ullrich2012DynamicalCM}
Paul~A. Ullrich, Christiane Jablonowski, James Kent, Peter~H. Lauritzen, and Ramachandran~D. Nair.
\newblock Dynamical core model intercomparison project ( {DCMIP} ) test case document.
\newblock 2012.
\newblock URL \url{https://api.semanticscholar.org/CorpusID:197412504}.

\bibitem[Ullrich et~al.(2018)Ullrich, Reynolds, Guerra, and Taylor]{Ullrich2018}
Paul~A. Ullrich, Daniel~R. Reynolds, Jorge~E. Guerra, and Mark~A. Taylor.
\newblock Impact and importance of hyperdiffusion on the spectral element method: A linear dispersion analysis.
\newblock \emph{Journal of Computational Physics}, 375:\penalty0 427--446, 2018.
\newblock \doi{10.1016/j.jcp.2018.06.035}.

\bibitem[Volpe(1993)]{volpe1993}
G.~Volpe.
\newblock Performance of compressible flow codes at low mach numbers.
\newblock \emph{AIAA Journal}, 31\penalty0 (1):\penalty0 49--56, 1993.
\newblock \doi{10.2514/3.11317}.
\newblock URL \url{https://doi.org/10.2514/3.11317}.

\bibitem[Weiss and Smith(1995)]{weiss1995}
Jonathan~M. Weiss and Wayne~A. Smith.
\newblock Preconditioning applied to variable and constant density flows.
\newblock \emph{AIAA Journal}, 33\penalty0 (11):\penalty0 2050--2057, 1995.
\newblock \doi{10.2514/3.12946}.
\newblock URL \url{https://doi.org/10.2514/3.12946}.

\bibitem[Witherden and Vincent(2020)]{witherden2020nodal}
Freddie Witherden and Peter Vincent.
\newblock On nodal point sets for flux reconstruction.
\newblock \emph{Journal of Computational and Applied Mathematics}, page 113014, 2020.

\bibitem[Yu and Jan(2020)]{YuJan2020}
Jian Yu and Hesthaven~S. Jan.
\newblock A study of several artificial viscosity models within the discontinuous galerkin framework.
\newblock \emph{Communications in Computational Physics}, 27\penalty0 (5):\penalty0 1309--1343, 2020.
\newblock \doi{10.4208/cicp.OA-2019-0118}.

\end{thebibliography}

\end{document}